\documentstyle[12pt, epsf]{article}

\renewcommand{\thefootnote}{\fnsymbol{footnote}}

\def\appendix#1{
  \addtocounter{section}{1}
  \setcounter{equation}{0}
  \renewcommand{\thesection}{\Alph{section}}
  \section*{Appendix}
  \addcontentsline{toc}{section}{Appendix \thesection\ \ \ #1}
  }

\newcommand{\newsection}{    % Numeration of eqs. is automatic
\setcounter{equation}{0}
\section}

\def \ov {\over }

\def\bea{\begin{eqnarray}}
\def\eea{\end{eqnarray}}

\def\LB{\left(}
\def\RB{\right)}
\def\be{\begin{equation}}
\def\ee{\end{equation}}

\def \bi{\bibitem}

\def\C{{\cal C}}
\def\tcm{\mbox{${\tilde{\cal M}}$}}

\def\te{\mbox{${\tilde{E}}$}}
\def\tcn{\mbox{${\tilde{\cal N}}$}}
\def\tp{\mbox{${\tilde P}$}}

\def\e9{\mbox{${E_9}$}}
\def\tph{{{\tilde \phi}}}
\def\ph{{{\phi}}}
\def\tf{\mbox{${\tilde F}$}}
\def\tt{\mbox{${\tilde T}$}}
\def\tl{\mbox{${\tilde \Lambda}$}}
\def\to{\mbox{${\tilde \Omega}$}}

\def \t{{\cal T}}
\def\es{\mbox{${E\llap/}$}}
\def\o{\mbox{${\Omega}$}}

\font\mybb=msbm10 at 12pt

\def\bb#1{\hbox{\mybb#1}}

\def\Z {\bb{Z}}
\def\T {\bb{T}}
\def\id{\protect{{1 \kern-.28em {\rm l}}}}

\begin{document}

\setcounter{page}{1}
\renewcommand{\thefootnote}{\arabic{footnote}}
\setcounter{footnote}{0}

\begin{titlepage}
\begin{flushright}
ITP-SB-00-80\\
\end{flushright}
\vspace{1cm}

\begin{center}
{\LARGE T-duality of the Green-Schwarz superstring}

\vspace{1.1cm}
{\large Bogdan Kulik${}$\footnote{
E-mail: bkulik@insti.physics.sunysb.edu }\,\,\,\,\,\,\,\,
Radu Roiban${}$\footnote{
E-mail: roiban@insti.physics.sunysb.edu } }\\

\vspace{18pt}

{\it C.N. Yang Institute for Theoretical Physics}

{\it SUNY at Stony Brook}

{\it  NY11794-3840, USA}
\\
\end{center}
\vskip 0.6 cm

\begin{abstract}
We study T-duality in the Green-Schwarz formalism to all orders in 
superspace coordinates. We find two analogs of Buscher rules for
the supervielbein and clarify their meaning from the superstring 
point of view. The transformation rules for the dilaton,
spin 1/2 fermions and  Ramond-Ramond superfields are also derived. 
\end{abstract}
\end{titlepage}

\newsection{Introduction}
Since its discovery, T-duality has been studied from 
various points of view. Together with the other duality symmetries, 
it has been the source of many recent developments in string theory. 
Being based on manipulations that do not 
change the corresponding conformal field theory, T-duality is probably 
the most securely founded. First identified in the sigma-model \cite{busher} 
and at the level of perturbative string spectra \cite{jap}, it is believed 
to be an exact symmetry of string theory. Moreover, it has been 
shown \cite{BHO}, \cite{BEK} that the low energy effective actions 
of type II string theories posses this symmetry. 

The action of T-duality on the NS-NS sector of string theory
can be constructed rather easily in the NSR formalism. However,
the proper treatment of the RR potentials in this formalism is
still unknown. Therefore, one has to find alternative approaches
to study their transformation under T-duality. 
One possible approach is to study the vertex operators associated 
with the RR fields and from there to infer their transformations.
This has been considered in \cite{dWLN} and \cite{pol}. In \cite{dWLN}
the authors have also considered the 9-dimensional 
spectrum of BPS states realized as wrapped $M2$ branes. The analysis 
is based on $N=2,\,D=9$ supersymmetry algebra. The results
give further support for the existence of a M-theory/IIB duality
\cite{JHS} that can be understood in terms of a fundamental supermembrane.

Another approach, considered in \cite{BHO}, \cite{BEK}, 
consists in studying the low energy effective actions of type II 
string theories, IIA/B supergravities, and constructing the explicit 
map between their field content. Since the supergravity theories
contain informations about world sheet 
quantum corrections, it necessarily
takes into account the quantum corrected T-duality, that is
the dilaton shift. This approach considers all orders in superspace 
coordinates.
A hybrid of the previous two approaches was considered 
in \cite{H1}, \cite{H2}. There, the author considers the action 
of T-duality on space-time spinors (gravitini, dilatini and 
supersymmetry parameters) and from there infers the transformation 
rules for RR fields.
Yet another approach considers the Green-Schwarz superstring up to some
given order in anticommuting superspace coordinates
\cite{CLPS}. Here the authors considered terms up to second
order in $\theta$ in the IIA superstring theory. Such terms 
involve the R-R field couplings. Their
transformation under T-duality has been constructed and the
IIB action up to ${\cal O}(\theta^2)$ has been written down.
T-duality in  massive type II theories has also been analyzed.
Duality to all orders in the odd superspace coordinates for the 
heterotic string was considered in \cite{WS}.

In this note we study T-duality in the Green-Schwarz 
formalism, to all orders in superspace coordinates. 
In the following section we consider the Green-Schwarz superstring in flat 
space and construct in detail the dual theory. We study both the supersymmetry 
transformations as well as the Siegel symmetry and find the map 
between the corresponding parameters in the two theories. This will 
prove to be a useful guide in studying the Green-Schwarz superstring in
a curved background, which we consider in the next section. 
Using the Siegel 
symmetry and super-covariance requirements we construct the
action of T-duality on the supervielbein components carrying flat bosonic 
indices. By additionally requiring that T-duality is an involution, 
we obtain the Buscher rules for the full supervielbein
up to an arbitrary function. By this time the Siegel symmetry transformations
will have the right form to imply the conventional constraints as well as the 
other lower dimensional constraints of the corresponding supergravity theory.
On the other hand, these constraints can be explicitly computed using the dual 
fields. Requiring that they are indeed satisfied fixes the arbitrary function 
mentioned above. These constraints together with the Bianchi identities 
imply \cite{gates} all the other supergravity constraints. By explicitly 
computing them in terms of the dual fields we find the T-duality action on the 
dilaton and axion and their superparteners as well as the T-duality action on 
the RR field strengths. This will be done in section {\bf 4}.

Our result will be that the supervielbein and super two-form
the fields of the two type II theories are related by T-duality in the 
following way: (the other transformations are listed in section {\bf 4})

\noindent
$\bullet 1$ if the duality transformation changes the chirality of the second
space-time gravitino then
\bea
{{\te}_{(+) 9}}{}^a \!\!\!\!&=&\!\!\!\!
{1\ov G_{99}}{E_9}{}^a\,\,\,\,\,;\,\,\,\,\,\,\,\,\,\,\,\,\,\,\,
\,\,\,\,\,\,\,\,\,\,\,\,\,\,\,\,\,\,\,\,
\,\,\,\,\,\,\,\,
{{\te}_{(+){\hat M}}}{}^a ={E}_{\hat M}{}^a - 
{G_{{\hat M}9}\ov G_{99}}{E_9}{}^a+
{{\cal B}_{{\hat M}9}\ov G_{99}}E_9{}^a\nonumber\\
{\te_{(+)9}}{}^{1\alpha}\!\!\!\!&=&\!\!\!\!
-{{E_9}{}^{1\beta}\ov G_{99}}{(\Gamma_-)_\beta}{}^\alpha
\,\,\,\,\,;\,\,\,\,\,\,\,\,\,\,\,\,\,\,\,\,\,\,\,\,\,\,\,\,\,
{\te_{(+)9}}{}^{2\alpha}=
{{E_9}{}^{2\beta}\ov G_{99}}{(E_9{}^a\Gamma{}^a\Gamma_-)_\beta}{}^\alpha
\nonumber\\
{{\te}_{(+){\hat M}}}{}^{1\alpha} \!\!\!\!&=&\!\!\!\!({E}_{\hat M}{}^{1\beta} - 
{G_{{\hat M}9}\ov G_{99}}{E_9}{}{}^{1\beta}-
{{\cal B}_{{\hat M}9}\ov G_{99}}E_9{}^{1\beta}){(\Gamma_-)_\beta}{}^\alpha
\nonumber\\
{{\te}_{(+){\hat M}}}{}^{2\alpha} \!\!\!\!&=&\!\!\!\!({E}_{\hat M}{}^{2\beta} - 
{G_{{\hat M}9}\ov G_{99}}{E_9}{}^{2\beta}+
{{\cal B}_{{\hat M}9}\ov G_{99}}E_9{}^{2\beta})
{({E_9}{}^a\Gamma{}^a\Gamma_-)_\beta}{}^\alpha\nonumber\\
{\tilde {\cal B}}_{(+)9{\hat M}}\!\!\!\!&=&\!\!\!\!
-{\tilde {\cal B}}_{{(+)\hat M}9}=
-{ G_{9{\hat M}} \ov G_{99}} 
\,\,\,\,\,\,;\,\,\,\,\,\,\,\,\,\,\,\,
{\tilde{\cal B}}_{(+){\hat M}{\hat N}} = {{\cal B}}_{{\hat M}{\hat N}}+ 
{1\ov G_{99}} \big[{\cal B}_{9{\hat M}}G_{9{\hat N}} - 
G_{9{\hat M}}{\cal B}_{9{\hat N}}\big]
\label{eq:finalrez}
\eea

\noindent
$\bullet 2$ 
if the duality transformation changes the chirality of the first 
space-time gravitino then
\bea
{{\te}_{(-) 9}}{}^a \!\!\!\!&=&\!\!\!\!
{1\ov G_{99}}E_9{}^a\,\,\,\,\,;\,\,\,\,\,\,\,\,\,\,\,\,\,\,\,\,\,\,\,
\,\,\,\,\,\,\,\,\,\,\,\,
\,\,\,\,\,\,\,\,
{{\te}_{(-){\hat M}}}{}^a ={E}_{\hat M}{}^a - 
{G_{{\hat M}9}\ov G_{99}}{E_9}{}^a-
{{\cal B}_{{\hat M}9}\ov G_{99}}E_9{}^a\nonumber\\
{\te_{(-)9}}{}^{1\alpha}\!\!\!\!&=&\!\!\!\!
{{E_9}{}^{1\beta}\ov G_{99}}{(E_9{}^a\Gamma{}^a\Gamma_+)_\beta}{}^\alpha
\,\,\,\,\,;\,\,\,\,\,\,\,\,\,\,\,\,\,
{\te_{(- )9}}{}^{2\alpha}=-
{{E_9}{}^{2\beta}\ov G_{99}}{(\Gamma_+)_\beta}{}^\alpha\nonumber\\
{{\te}_{(-){\hat M}}}{}^{1\alpha} \!\!\!\!&=&\!\!\!\!({E}_{\hat M}{}^{1\beta} - 
{G_{{\hat M}9}\ov G_{99}}{E_9}{}^{1\beta}-
{{\cal B}_{{\hat M}9}\ov G_{99}}E_9{}^{1\beta})
{({E_9}{}^a\Gamma{}^a\Gamma_+)_\beta}{}^\alpha\nonumber\\
{{\te}_{(-){\hat M}}}{}^{2\alpha} \!\!\!\!&=&\!\!\!\!({E}_{\hat M}{}^{2\beta} - 
{G_{{\hat M}9}\ov G_{99}}{E_9}{}^{2\beta}+
{{\cal B}_{{\hat M}9}\ov G_{99}}E_9{}^{2\beta}){(\Gamma_+)_\beta}{}^\alpha
\nonumber\\
{\tilde {\cal B}}_{(-)9{\hat M}}\!\!\!\!&=&\!\!\!\!
-{\tilde {\cal B}}_{{(-)\hat M}9}=
{G_{9{\hat M}}\ov G_{99}}
\,\,\,\,\,\,;\,\,\,\,\,\,\,\,\,\,\,\,\,\,
{\tilde{\cal B}}_{(-){\hat M}{\hat N}} = {{\cal B}}_{{\hat M}{\hat N}}+ 
{1\ov G_{99}} \big[{\cal B}_{9{\hat M}}G_{9{\hat N}} - 
G_{9{\hat M}}{\cal B}_{9{\hat N}}\big]
\label{eq:finalrez1}
\eea
In the above formula the indices $1\alpha$ and $2\alpha$ are the two possible 
types of spinor indices according to whether the theory is type IIA/B,
and $G_{MN}$ is defined as ${E_M}{}^a {E_N}{}^b\eta_{ab}$.
These are the Buscher rules at the level of supervielbein
super-two forms are related by the usual rules. In all our 
manipulation we absorb the dilaton superfield by rescaling the
super-vielbeine. In this setup
the dilaton transformation is already 
encoded in (\ref{eq:finalrez}-\ref{eq:finalrez1}), as noted 
also in \cite{CLPS}. We will rederive it in section {\bf 4} from the 
matching of constraints. 
We will end with some comments and conclusions.

\newsection{T-duality in flat space}
Let us consider first the Green-Schwarz superstring 
in flat space, with the action given by:
\bea
S=&-&{1\ov 2\pi}\int d\tau d\sigma \sqrt{-h}\Pi_i{}^\mu 
\Pi_j{}^\nu h{}^{ij}\eta_{\mu\nu}-\nonumber\\
&-&{1\ov \pi}\int d\tau d\sigma \epsilon{}^{ij}
[\partial_i X{}^\mu (\bar\theta{}^1\Gamma{}^\nu\partial_j\theta{}^1
-\bar\theta{}^2\Gamma{}^\nu\partial_j\theta{}^2)+
\bar\theta{}^1\Gamma{}^\nu\partial_i\theta{}^1
\bar\theta{}^2\Gamma{}^\nu\partial_j\theta{}^2]\eta_{\mu\nu}\,\,\,\,.
\eea
with the supervielbein pullback
\be
{\Pi}_i{}^\mu=\partial_i X{}^\mu - 
\bar\theta{}^1\Gamma{}^\mu\partial_i\theta{}^1
-\bar\theta{}^2\Gamma{}^\mu\partial_i\theta{}^2\,\,\,\,.
\ee
We proceed in the standard way (\cite{RV}, \cite{busher}),
set $A_i=\partial_i X{}^9$ and introduce a Lagrange multiplier 
${\tilde X}$ to enforce the constraint that $A_i$ is flat.
We split the action $S$ 
into $S_9(A_i)$ and ${\hat S}$ independent of $A_i$ and add
\be
S_{aux}={1\ov \pi}\int d\tau d\sigma {\tilde X}\epsilon{}^{ij}
\partial_i A_j\,\,\,\,.
\ee
By requiring supersymmetry invariance of the modified action and 
keeping in mind that $\delta A_i=\partial_i\delta X{}^9$ one gets the
supersymmetry variation of the 
Lagrange multiplier ${\tilde X}$. Obviously, $\delta_{susy}{\hat S}=0$. 
Thus, the variation of the Lagrange multiplier comes 
from $\delta_{susy} S_9$  
and it reads:
\be
\delta {\tilde X}=\bar\epsilon{}^1\Gamma{}^\nu\theta{}^1-\bar\epsilon{}^2
\Gamma{}^\nu\theta{}^2\,\,\,\,.
\ee

To construct the dual action we integrate out $A_i$; at the classical 
level we use its field equation. Since we are in flat space, there are 
no quantum corrections at one loop level. The resulting action is: 
\bea
{\tilde S}=&-&{1\ov 2\pi}\int d\tau d\sigma\sqrt{-h} 
{\tilde \Pi}_i{}^\mu 
{\tilde\Pi}_j{}^\nu
h{}^{ij}\eta_{\mu\nu}-\nonumber\\
&-&{1\ov \pi }\int d\tau d\sigma \epsilon{}^{ij}
[\partial_i  {\check{X}}{}^\mu
(\bar\theta{}^1\Gamma{}^\nu\partial_j\theta{}^1
-(-){}^{\delta_{\nu 9}}\bar\theta{}^2\Gamma{}^\nu\partial_j\theta{}^2)
+\nonumber\\
&+&(-){}^{\delta_{\nu 9}}
\bar\theta{}^1\Gamma{}^\nu\partial_i\theta{}^1
\bar\theta{}^2\Gamma{}^\nu\partial_j\theta{}^2]
\eta_{\mu\nu}
\eea
\be
{\tilde \Pi}_i{}^\nu=(\eta){}^{\delta_{\nu 9}}\big(\partial_i 
{\check{X}}{}^\nu - \bar\theta{}^1\Gamma{}^\nu\partial_i\theta{}^1
-(-){}^{\delta_{\nu 9}}\bar\theta{}^2\Gamma{}^\nu\partial_i\theta{}^2\big)
\ee
where ${\check{X}}{}^\nu = (X{}^{\hat\nu}, {\tilde X})$ and we 
introduce $\eta=\pm 1$.

Requiring covariance for ${\tilde \Pi}_i{}^\nu$ expressed in terms of the
dual variables, we are led to the following redefinitions:
\be
\eta=+1:\,\,\,\,\left\{\begin{array}{rrrccc}
{\tilde X}{}^\mu\!\!\!&=&\!\!\!(X{}^{\hat \mu},\,{\tilde X})\\
{\tilde\theta}{}^1\!\!\!&=&\!\!\! 
{\theta}{}^1{\,\,\,\,\,\,\,\,\,\,\,\,\,\,\,\,}\\
{\tilde\theta}{}^2\!\!\!&=&\!\!\!\Gamma{}^9 {\theta}{}^2{\,\,\,\,\,\,\,\,\,\,}
\\\end{array}\right.
\,\,\,\,\,\,\,\,\,\,\eta=-1:\,\,\,\,\left\{\begin{array}{rrrccc}
{\tilde X}{}^\mu\!\!\!&=&\!\!\!(X{}^{\hat \mu},\,-{\tilde X})\\
{\tilde\theta}{}^1\!\!\!&=&\!\!\! \Gamma{}^9{\theta}{}^1{\,\,\,\,\,\,\,\,\,\,
\,\,\,\,}\\
{\tilde\theta}{}^2\!\!\!&=&\!\!\! {\theta}{}^2{\,\,\,\,\,\,\,\,\,\,\,\,\,\,
\,\,\,\,\,\,}
\\\end{array}\right.\,\,\,\,.
\label{eq:redefflat}
\ee
In terms of these variables ${\tilde \Pi}_i{}^\mu$ is covariant and
has the same
form as ${\Pi}_i{}^\mu$. Therefore, the dual action has the same 
form as the original one. These 
transformations flipped the chirality of one of the 
spinor variables, thus interchanging the two type II theories in 
flat space.

Let us now consider the Siegel symmetry transformation of the 
initial action and its dual. Following the obvious statement that 
the variation of a Lagrangian is a sum of variations of fields times 
their equations of motion, we first write 
down these equations. Hatted indices run from 0 to 8.
\bea
0=\pi {\delta L\ov\delta X{}^{\hat\mu}}=\partial_i(\sqrt{-h}
h{}^{ij}\Pi_j{}^{\hat\mu})+\epsilon{}^{ij}(
\partial_i{\bar\theta}{}^1\Gamma{}^{\hat\mu}\partial_j\theta{}^1-
\partial_i{\bar\theta}{}^2\Gamma{}^{\hat\mu}\partial_j\theta{}^2)\,\,\,\,,
\nonumber
\eea
\be
0=\pi {\delta L\ov\delta A_i}=\epsilon{}^{ij}
\partial_j{\tilde X} -\sqrt{-h}h{}^{ij}\Pi_j{}^9-
\epsilon{}^{ij}(
\partial_i{\bar\theta}{}^1\Gamma{}^9\partial_j\theta{}^1-
\partial_i{\bar\theta}{}^2\Gamma{}^9\partial_j\theta{}^2)\,\,\,\,,
\ee
\bea
0=-2\pi {\delta L\ov\delta {\bar\theta}{}^1}=-2\pi\Gamma{}^{\hat\mu}\theta{}^1
\Big({\delta L\ov\delta X{}^{\hat\mu}}\Big)&+&2\pi\Gamma{}^9\theta{}^1
\partial_i\Big({\delta L\ov\delta A_i}\Big)-
2\epsilon{}^{ij}\partial_i A_j\Gamma{}^9\theta{}^1
-\nonumber\\
&-&8\sqrt{-h}P_{-}{}^{ij}\big(\Pi_i{}^{\hat\mu}\Gamma_{\hat\mu}
\partial_j\theta{}^1+\Pi_i{}^9\Gamma_9
\partial_j\theta{}^1\big)\,\,\,\,,\nonumber
\eea
\bea
0=-2\pi {\delta L\ov\delta {\bar\theta}{}^2}=-2\pi\Gamma{}^{\hat\mu}\theta{}^2
\Big({\delta L\ov\delta X{}^{\hat\mu}}\Big)&+&2\pi\Gamma{}^9\theta{}^2
\partial_i\Big({\delta L\ov\delta A_i}\Big)-
2\epsilon{}^{ij}\partial_i A_i\Gamma{}^9\theta{}^2
-\nonumber\\
&-&8\sqrt{-h}P_{+}{}^{ij}\big(\Pi_i{}^{\hat\mu}\Gamma_{\hat\mu}
\partial_j\theta{}^2+\Pi_i{}^9\Gamma_9
\partial_j\theta{}^2\big)\,\,\,\,.\nonumber
\eea

Using the Siegel transformations of $X{}^{\hat \mu}$
\be
\delta_\kappa X{}^{\hat\mu}=-\delta_\kappa{\bar\theta}{}^I
\Gamma{}^{\hat\mu}\theta{}^I
\label{eq:kxh}
\ee
as well as the variation of the vector field $A_i$  
\be
\delta_\kappa A_i=-\partial_i\big(
\delta_\kappa\bar\theta{}^I\Gamma{}^9\theta{}^I\big)
\ee
we get the variation of the Lagrange multiplier:
\be
\delta_\kappa{\tilde X}=\eta\delta_\kappa{\tilde X}{}^9
=-\delta_\kappa{\bar\theta}{}^1\Gamma{}^9\theta{}^1+
\delta_\kappa{\bar\theta}{}^2\Gamma{}^9\theta{}^2\,\,\,\,.
\label{eq:kxt}
\ee
We also have to express the variation of the spinor coordinates in 
terms of the dual variables. Originally we had:
\be
\delta_\kappa\theta{}^I=\Pi_i{}^{\hat\mu}\Gamma_{\hat\mu}
\kappa{}^{i I}
+\Pi_i{}^9\Gamma_9\kappa{}^{i I}\,\,\,\,.
\ee
Using $A_i$ field equation one can express $\Pi_i{}^9$
in terms of the dual variables as follows:
\be
\Pi_i{}^9=h_{ij}{\epsilon{}^{jl}\ov\sqrt{-h}}
[\partial_l{\tilde X}-({\bar\theta}{}^1\Gamma{}^9\partial_l\theta{}^1-
{\bar\theta}{}^1\Gamma{}^9\partial_l\theta{}^1)]=
\eta h_{ij}{\epsilon{}^{jl}\ov\sqrt{-h}}
{\tilde\Pi}_l{}^9\,\,\,\,.
\ee
Using the constraints on the parameters $\kappa{}^{1i}$ and $\kappa{}^{2i}$
we can write
\be
\Pi_i{}^9\Gamma_9\kappa{}^{i 1}=-\eta {\tilde \Pi}_j{}^9\Gamma_9
{\epsilon{}^{jl}\ov\sqrt{-h}}h_{lm}\kappa{}^{m 1}=\eta
{\tilde \Pi}_j{}^9\Gamma_9\kappa{}^{j 1}
\label{eq:k1}
\ee
and
\be
\Pi_i{}^9\Gamma_9\kappa{}^{i 2}=-\eta {\tilde \Pi}_j{}^9\Gamma_9
{\epsilon{}^{jl}\ov\sqrt{-h}}h_{lm}\kappa{}^{m 2}=-\eta
{\tilde \Pi}_j{}^9\Gamma_9\kappa{}^{j 2}\,\,\,\,.
\label{eq:k2}
\ee
Combining (\ref{eq:k1}) and (\ref{eq:k2}) together with (\ref{eq:redefflat}),
(\ref{eq:kxh}) and (\ref{eq:kxt}), one arrives at 
the conclusion that the parameter of the Siegel transformation 
transforms under T-duality as follows:
\be
\kappa{}^{i 2}\longrightarrow 
{\tilde\kappa}{}^{i 2}=-
%%%{1\ov R} 
\Gamma{}^9\kappa{}^{i 2}\,\,\,\,,\,\,\,\,\,\,\,\,\,\,\,\,
\,\,\,\,\,\,\,\,
\kappa{}^{i 1}\longrightarrow 
{\tilde\kappa}{}^{i 1}=-
%%%{1\ov R} 
\Gamma{}^9\kappa{}^{i 1}
\ee
for $\eta=+1$ and $\eta=-1$, respectively.
Obviously, all its properties are the same as in the initial theory
except its chirality which was flipped. The redefinitions found above
will serve as a guide in considering type II superstring in arbitrary 
supergravity background.

\newsection{Arbitrary supergravity background}
We now turn to the type II theories in arbitrary supergravity 
background. Since the explicit form
of supervielbein and NS-NS super-two-form are not known to all 
orders in the odd superspace coordinate,
one has to use the superspace constraints to show that under 
T-duality the type II theories are mapped into each other.
The same procedure allows us to construct the Buscher rules
at the level of supervielbein. These rules, together
with those for the NS-NS super-two-form, implicitely contain
the T-duality transformations for all fields in the theory.

It is well known [10, 11] that the Green-Schwarz superstring 
is invariant under the Siegel transformations if the 
supervielbein and the super-two-form satisfy the
supergravity constraints. At the same time however,
invariance under Siegel transformations only requires
that certain linear combinations of supergravity 
constraints hold.
Therefore, our strategy will be the following:
first we study the Siegel transformations in the dual theory;
Combining them with covariance requirements we will construct four
possible vielbein solutions of the Buscher rules for the 
``super-metric''; Requiring further that T-duality is an involution
we will discard two of the possible vielbein with flat bosonic index
and we will show that the vielbein with flat spinor indices can be 
redefined such that the Siegel transformations have the standard form.
This determines all the components of the supervielbein up to 
an arbitrary function. This function will be determined 
by explicitly computing the constraints required by  the Siegel transformations 
in terms of the dual fields and requiring that they are satisfied.
The other constraints follow from Bianchi identities as shown in \cite{gates}.
We will use them to derive the transformation of fields that do not appear 
explicitly in the $\sigma$ model.

Consider the type IIA action in a general supergravity 
background, which can be obtained
by dimensional reduction of 11-dimensional supermembrane action 
\cite{DHIS}, and assume that all fields are independent on the coordinate
$X{}^9$. Replacing $\partial_i X{}^9$ by the vector field $A_i$ and
constraining it to be flat we get the action:
\bea
S&=&\int d{}^2\xi\big(\,{1\ov 2}\Phi \sqrt{-g}g{}^{ij}E_i{}^a E_j{}^b\eta_{ab}+
\epsilon{}^{ij}\partial_iZ{}^M\partial_jZ{}^N {\cal B}_{NM} + 
\epsilon{}^{ij}{\tilde X}\partial_iA_j\big)\,\,\,\, ,\\
{\cal B}_{MN}&=&(-){}^{M(N+A)}E_N{}^A E_M{}^B{\cal B}_{BA}
\,\,\,\,\,\,\,\,{\rm and}\,\,\,\,\,\,\,\,
E_i{}^a\equiv A_i {E_9}{}^a+\partial_i Z{}^{\hat M}{E_{\hat M}}{}^a\,\,\,\,.
\nonumber
\eea 
The dilaton superfield $\Phi$ can be removed by a suitable rescaling 
of the supervielbein, so we will set it equal to 1 in the following.

Defining:
\be
G_{99}=E_9{}^a E_9{}^b\eta_{ab} \,\, , \hskip1truecm
\hskip1.2truecm G_{9{\hat M}}=
E_9{}^a E_{\hat M}{}^b\eta_{ab} \,\, ,\hskip1truecm\hskip1.2truecm
 G_{{\hat M}{\hat N}}=
E_{\hat M}{}^{a} E_{\hat N}{}^b\eta_{ab}\,\,\,\,,
\ee
integrating out $A_i$ using its field equation
\bea
E_i{}^a&=&\partial_iZ{}^{\hat M}\big({E_{\hat M}}{}^a-
{G_{{\hat M}9}\ov G_{99}}{E_9}{}^a\big)-
{g_{ij}\epsilon{}^{jk}\ov\sqrt{-g}G_{99}}\big(\partial_k{\tilde X}+
\partial_kZ{}^{\hat M}{\cal B}_{{\hat M}9}\big){E_9}{}^a
\label{eq:Afeq}
\eea
and allowing for the identifications ${\tilde X}{}^9=\eta{\tilde X}$ with
$\eta=\pm1$, we get:

-dual ``supermetric'':
\bea
{\tilde G}_{99}&=&{1\ov G_{99}}\nonumber\\
{\tilde G}_{9{\hat M}}&=&-{\eta\ov G_{99}} {\cal B}_{9{\hat M}}=
{\tilde G}_{{\hat M}9}\nonumber\\
{\tilde G}_{{\hat M}{\hat N}} &=& {G}_{{\hat M}{\hat N}}- {1\ov G_{99}}
\big[G_{9{\hat M}}G_{9{\hat N}} - 
{\cal B}_{9{\hat M}}{\cal B}_{9{\hat N}}\big]
\label{eq:GMN}
\eea

-dual super 2-form:
\bea
{\tilde {\cal B}}_{9{\hat M}}&=&-{\eta\ov G_{99}} G_{9{\hat M}}
=-{\tilde {\cal B}}_{{\hat M}9}\nonumber\\
{\tilde{\cal B}}_{{\hat M}{\hat N}} &=& {{\cal B}}_{{\hat M}{\hat N}}+ 
{1\ov G_{99}} \big[{\cal B}_{9{\hat M}}G_{9{\hat N}} - 
G_{9{\hat M}}{\cal B}_{9{\hat N}}\big]\,\,\,\,\,\,.
\label{eq:BMN}
\eea

Following the same steps as in flat space and using the Siegel
transformations of the IIA theory \cite{BST} 
we deduce the 
Siegel transformation of the ${\tilde X}{}^9$ coordinate:
\be
\delta_\kappa {\tilde X}{}^9=-\eta\delta_\kappa Z{}^{\hat M}
 {\cal B}_{{\hat M}9}\,\,\,\, .\label{eq:lmk}
\ee

Let us now check whether the dualized variables obey the type IIB 
Siegel transformation rules. First we notice that using the Siegel 
symmetry of type IIA theory, $\delta_\kappa E{}^a\equiv
\delta_\kappa Z{}^M E_M{}^a=0$, 
 one can write:
\be
\delta_\kappa Z{}^M G_{M{N}}=0\,\,\,\,\Longrightarrow\,\,\,\,
\delta_\kappa Z{}^{\hat M} G_{{\hat M}{N}}=-\delta_\kappa X{}^9 
G_{9{N}}
\label{eq:idkappa}
\ee
which together with \ref{eq:BMN} and \ref{eq:lmk} leads to
\bea
\!\!\!\!
\delta_\kappa {\tilde Z}^N {\te}_N{}^a 
{\te}_{\hat M}{}^b \eta_{ab}&=&
\delta_\kappa {\tilde X}^9{\tilde G}_{9{\hat M}} +
\delta_\kappa Z^{\hat N}{\tilde G}_{{\hat N}{\hat M}}=\nonumber\\
&=&-\eta\delta_\kappa Z^{\hat N}{\cal B}_{{\hat N}9}
{\eta\ov G_{99}}{\cal B}_{{\hat M}9}+
\delta_\kappa Z^{\hat N}\big[{G}_{{\hat N}{\hat M}}- {1\ov G_{99}}
(G_{9{\hat N}}G_{9{\hat M}} - 
{\cal B}_{9{\hat N}}{\cal B}_{9{\hat M}})\big]=\nonumber\\
&=&\delta_\kappa Z{}^{\hat N}\big[{G}_{{\hat N}{\hat M}}- 
{1\ov G_{99}}G_{9{\hat N}}G_{9{\hat M}}\big]\,\,\,\,\,\,.
\eea
The contraction with ${E_9}{}^a$ must be considered separately and 
we obtain
\be
\delta_\kappa {\tilde Z}{}^N {\te}_N{}^a {\te}_{9}{}^b \eta_{ab}
=-\eta\delta_\kappa Z{}^{\hat M}{\cal B}_{{\hat M}9}{\tilde G}_{99}+
\delta_\kappa Z{}^{\hat M}{\tilde G}_{{\hat M}9}=0\,\,\,\,\,\,.
\label{eq:KB2}
\ee
Multiplying by ${\te_a}{}^M$, we find
\be
\delta_\kappa {\tilde Z}{}^N {\te}_N{}^a=0\,\,\,\,.
\label{eq:first}
\ee
Notice that in deriving this result we did not need to know
the action of T-duality at the level of supervielbein. This is, however,
not a generic feature as we will see by considering the spinor 
part of Siegel transformations.

Start from the IIA theory and introduce two 
vielbeine, ${E_M}{}^{I\alpha}, I=1,2$, subject to 
chirality constraints 
${{E}_M}{}^{1\alpha}{(\Gamma_+)_\alpha}{}^\beta=0$ and 
${{E}_M}{}^{2\alpha}{(\Gamma_-)_\alpha}{}^\beta=0$. 
Here $\alpha$ denotes a 10-dimensional spinor index,
$\alpha\equiv ({\dot\alpha},\,\alpha)$ and 
$\Gamma_\pm={1\ov 2}(1\pm\Gamma_{11})$.
The Siegel symmetry 
transformations of IIA theory are given by \cite{DHIS}:
\be
\delta_\kappa E{}^{1\alpha}= E_i{}^aP_+{}^{ij} 
\kappa_j{}^{1\rho}{(\Gamma_+)_\rho}{}^\beta
{(\Gamma_a)_\beta}{}^\alpha\,\,\,\,\,\,\,\,,\,\,\,\,\,\,\,\,
\delta_\kappa E{}^{2\alpha}= E_i{}^a 
P_-{}^{ij}\kappa_j{}^{2\rho}{(\Gamma_-)_\rho}{}^\beta
{(\Gamma_a)_\beta}{}^\alpha
\ee
with the usual 
$P_\pm{}^{ij}={1\ov 2}(g{}^{ij}\pm{\epsilon{}^{ij}\ov \sqrt{-g}})$
and 
$\delta_\kappa E{}^{I\alpha} \equiv \delta_\kappa Z{}^M {E_M}{}^{I\alpha}$.
Let us also introduce for the 
dual theory two types of vielbein with flat spinor indices
(${\te}{}^{I\alpha},\,\,I=1,2$). For each of the two vielbeine 
the Siegel variation is given by: 
\be
\delta_\kappa {\te}{}^{I\alpha}\equiv
\delta_\kappa {\tilde X}{}^9{{\te}_9}{}^{I\alpha}
+\delta_\kappa Z{}^{\hat M}{{\te}_{\hat M}}{}^{I\alpha}=\delta_\kappa 
Z{}^{\hat M}[
{{\te}_{\hat M}}{}^{I\alpha}-\eta
{\cal B}_{{\hat M}9}{{\te}_9}{}^{I\alpha} ]
=
\delta_\kappa Z{}^{\hat M}[
{{\te}_{\hat M}}{}^{I\alpha}-
{{\tilde G}_{{\hat M}9}\ov{\tilde G}_{99}}{{\te}_9}{}^{I\alpha} ]\,\,,
\ee
where we have used (\ref{eq:lmk}).
Let us introduce the notation:
\be
{\tilde P}_{\hat M}{}^{I\alpha}={\te}_{\hat M}{}^{I\alpha}-
{{\tilde G}_{{\hat M}9}\ov{\tilde G}_{99}}{\te}_9{}^{I\alpha}\,\,\,\,.
\ee
From the original theory we have:
\be
\delta_\kappa Z{}^{\hat M}=\delta_\kappa E{}^{I\alpha} {E_{I\alpha}}{}^{\hat M}
\,\,\,\,.\ee
Rewriting the $A_i$ field equation (\ref{eq:Afeq}) using the dual 
variables one gets for ${E_i}{}^a$\,:
\be
E_i{}^a=\partial_iZ{}^{\hat M}E_{\hat M}{}^a - \eta\big[\,g_{ij}
\big(P_+{}^{jk}-P_-{}^{jk}\big)\,\partial_k {\tilde Z}{}^{M} {\tilde G}_{M9}+
\partial_iZ{}^{\hat M}{\tilde{\cal B}}_{{\hat M}9}\big]\,E_9{}^a
\ee
Thus, recalling that $P_+{}^{ij} P_-{}^{kl} g_{jk} = 0 $ and
                     $P_+{}^{ij} P_+{}^{kl} g_{jk} = P_+{}^{il} $
one gets for the variations in the dual theory:
\bea
\delta_\kappa {\te}^{I\alpha}\!\!\!\!&=&\!\!\!\!
\Big\{\partial_iZ^{\hat M}\left(
E_{\hat M}{}^a-\eta {\tilde{\cal B}}_{{\hat M}9}E_9{}^a 
\right)
+\eta \partial_i {\tilde Z}^{M} {\tilde G}_{M9}
E_9{}^a\Big\}P_+^{ij}\kappa_j^{1\rho}
{(\Gamma_+)_\rho}{}^\beta{(\Gamma_a)_\beta}^\sigma
 {E_\sigma}^{1\hat M}
{\tilde P}_{\hat M}{}^{I\alpha}+\nonumber\\
&+&\!\!\!\!\Big\{\partial_i Z^{\hat M}
\left(
E_{\hat M}{}^a -\eta{\tilde{\cal B}}_{{\hat M}9}E_9{}^a 
\right)
- \eta\partial_i {\tilde Z}^{M} {\tilde G}_{M9}
E_9{}^a\Big\}P_-^{ij}\kappa_j^{2\rho}{(\Gamma_-)_\rho}^\beta
{(\Gamma_a)_\beta}^\sigma
{E_\sigma}^{2\hat M}
{{\tilde P}_{\hat M}}{}^{I\alpha}
\label{eq:spink}
\eea
From the above expressions, on covariance ground, we can make 
the following identification: 
\bea
{{\te}_{(\sigma,\eta) 9}}{}^a &=& \sigma\eta 
{1\ov G_{99}}E_9{}^a\,\,\,\,\,\,\,\,\,\,\,\,,
\nonumber\\
{{\te}_{(\sigma,\eta){\hat M}}}{}^a &=&{E}_{\hat M}{}^a - 
{G_{{\hat M}9}\ov G_{99}}{E_9}{}^a+
\sigma |\eta| {{\cal B}_{{\hat M}9}\ov G_{99}}E_9{}^a
\label{eq:solb} 
\eea
where $\sigma=\pm 1$ independently of $\eta$ and the absolute value 
of $\eta$ in the second equation is for notational convenience.
It is easy to check that (\ref{eq:solb}) are solutions for 
(\ref{eq:GMN}) for all $\sigma$ and $\eta$.
These equations represent the action of T-duality on the vielbein
components with flat bosonic indices. As one probably expects, using these 
expressions for the dual vielbein one can immediately get 
equation (\ref{eq:first}):
\bea
\delta_\kappa {\te_{(\sigma,\eta)}}{}^a&=&\delta_\kappa Z{}^{\hat M}\Big\{
-\sigma{\cal B}_{{\hat M}9}
{E_9{}^a\ov G_{99}}+E_{\hat M}{}^a - {G_{{\hat M}9}\ov G_{99}}E_9{}^a
+\sigma{{{\cal B}}_{{\hat M}9}\ov G_{99}}
E_9{}^a\Big\}\nonumber\\
&=&\delta_\kappa Z{}^{\hat M}\Big\{
E_{\hat M}{}^a-{G_{{\hat M}9}\ov G_{99}}E_9{}^a\Big\}=0
\eea
upon using (\ref{eq:idkappa}) for the last equal sign.
At this point we seem to have four possible choices of vielbein 
with flat bosonic indices, two for each choice of $\eta$.
This is twice as many as in \cite{BEK},\cite{H1},\cite{H2}, where similar
expressions have been considered. 

This is as much as we can get by considering only the Siegel transformations
and covariance. By requiring that T-duality is an involution we will obtain
additional useful informations. It is easy to check that 
equation (\ref{eq:solb}) satisfy the 
involution requirement for all values of $\sigma$ and $\eta$. 
However, 
we will be able to single out a unique vielbein for each choice of $\eta$
by studying the transformations of the $\kappa$ parameters.
In \cite{H2} it has been argued that both solutions are necessary for
the consistency of the theory. As we will see below, the two solutions
correspond to IIB theories with opposite chiralities.

The first step is to notice that (\ref{eq:spink}) can be cast in 
a more useful form using the following identity:
\be
\partial_\kappa{\tilde Z}{}^M {{\te}_{(\sigma,\eta) M}}{}^a\Gamma{}^a
=-\partial_\kappa{\tilde Z}{}^M{{\te}_{(-\sigma,\eta) M}}{}^a E\llap/_9
\Gamma{}^a E\llap/_9\,\,\,\,.
\ee
Here we have introduced the notation 
\be
{(E\llap/_9)_\alpha}{}^\beta\equiv 
{{E_9}{}^a\ov \sqrt{G_{99}}}{(\Gamma_a)_\alpha}{}^\beta\,\,\,\,,
\ee
which plays the role of $\Gamma_9$ in the flat case above.
Thus, with the redefinition 
\be
{\kappa}{}^{2\alpha}_j\longrightarrow {\tilde \kappa}{}^{2\alpha}_j=-
\kappa{}^{2\beta}_j{{E\llap/_9}_\beta}{}^\alpha
\,\,\,\,,\,\,\,\,\,\,\,\,\,\,\,\,\,\,\,\,
{\kappa}{}^{1\alpha}_j\longrightarrow {\tilde \kappa}{}^{1\alpha}_j=-
\kappa{}^{1\beta}_j{{E\llap/_9}_\beta}{}^\alpha
\label{eq:tdktr}
\ee
for $\sigma=+1$ and $\sigma=-1$ respectively, equation
(\ref{eq:spink}) can be written as follows:
\bea
\!\!\!\!\!\!\!\!\delta_\kappa {\te}{}^{I\alpha}\!\!\!\!&=&\!\!\!\!
\partial_i{\tilde Z}{}^N{{\te}_{(+,\eta)}}_N{}^a
\Big\{P_+{}^{ij}\kappa_j{}^{1\zeta}
%{(\Gamma_+)_\rho}{}^\zeta
{(\Gamma_a)_\zeta}{}^\beta E_{1\beta}{}^{\hat M}
 + P_-{}^{ij}
{\tilde\kappa}_j{}^{2\zeta}
%{(\Gamma_+)_\rho}{}^\zeta
{(\Gamma_a)_\zeta}{}^\sigma
{(E\llap/_9)_\sigma}{}^\beta E_{2\beta}{}^{\hat M}\Big\}
{\tilde P}_{\hat M}{}^{I\alpha}\nonumber\\
\!\!\!\!&=&\!\!\!\!\partial_i{\tilde Z}{}^N{{\te}_{(-,\eta)}}_N{}^a
\Big\{
P_+{}^{ij}{\tilde\kappa}_j{}^{1\zeta}
%{(\Gamma_-)_\rho}{}^\zeta
{(\Gamma_a)_\zeta}{}^\sigma
{(E\llap/_9)_\sigma}{}^\beta E_{1\beta}{}^{\hat M}
 + P_-{}^{ij}
\kappa_j{}^{2\zeta}
%{(\Gamma_-)_\rho}{}^\zeta
{(\Gamma_a)_\zeta}{}^\beta E_{2\beta}{}^{\hat M}
\Big\}
{\tilde P}_{\hat M}{}^{I\alpha}
\label{eq:second}
\eea
where we have absorbed the chiral projectors by letting them act on $\kappa$.
Notice that the redefinition 
of $\kappa{}^{1(2)}_j$ is the direct curved space analog of the flat 
space case. Notice also that equation (\ref{eq:tdktr}) can be 
written in a more compact form:
\be
{\tilde\kappa}_{(\eta)j}{}^{I\alpha}=
-{\kappa}_{(\eta)j}{}^{J\beta}{{{\cal T}_{(\eta)}}_{J\beta}}{}^{I\alpha}
\ee
with ${{{\cal T}_{(\eta)}}_{J\beta}}{}^{I\alpha}$ defined as follows:.
\be
{{\cal T}_{(+)}}_B{}^A=
\left\{\begin{array}{rrccc}\delta_B{}^A\,\,\,\,
&{\rm for}\,\,\,\,A,B\in&\!\!\!\!\{a,&\!\!\!1\alpha\}  \\
{(E\llap/_9)_\beta}{}^\alpha &{\rm for}\,\,\,\,A,B=&
\!\!\!\!2\alpha&\\\end{array}\right.
\,\,\,;\,\,\,
{{\cal T}_{(-)}}_B{}^A=
\left\{\begin{array}{rrccc}
\delta_B{}^A\,\,\,\,&{\rm for}\,\,\,\,A,B\in&
\!\!\!\!\{a,&\!\!\!2\alpha\}  \\
{(E\llap/_9)_\beta}{}^\alpha
 &{\rm for}\,\,\,\,A,B=&1\alpha&\\\end{array}\right.
\label{eq:deft}
\ee

{\it A priori} there is no principle allowing us to fix
${E_{I\beta}}{}^{\hat M} {\tilde P}_{\hat M}{}^{J\alpha}$ in terms of the 
fields of the IIA theory. The strategy for solving
this problem  will be the following: we 
parametrize the arbitrariness of the dual vielbein
via their projection on the original ones and see 
how much of it can be accounted for by field redefinitions that
do not change the form of the Lagrangian.
As stressed previously, we require that
the resulting Siegel variations are covariant and that T-duality is 
an involution up to relabeling of vielbein and  $\kappa$ parameters. 
The remaining pieces could be interpreted 
as the analog of Buscher rules for the vielbein with flat spinor 
indices.

Let us consider next the parameters of the Siegel transformations. For
${\te}{}^a={\te}_{(+,\eta)}{}^a$ we impose
that ${\tilde{\tilde\kappa}}{}^2=\kappa{}^2$ and we get 
$\eta =1$. For the other choice of dual 
vielbein, ${\te}{}^a={\te}_{(-,\eta)}{}^a$, 
we impose that ${\tilde{\tilde\kappa}}{}^1=\kappa{}^1$ and we get 
$\eta=-1$. These can be summarized in the following equation:
\be
\sigma\eta=1\,\,\,\,\Leftrightarrow\,\,\,\,\sigma=\eta
\ee
Thus, as promised, we are left with a unique vielbein with flat 
bosonic indices for each choice of $\eta$. 

Now we analyze the vielbein with flat spinor indices and show that 
the two values of $\eta$ 
correspond to the two IIB theories with opposite chiralities,
just as in the case of flat space. For notational convenience
we will replace the pair $(\sigma,\eta)$ by $\eta$. Since
${{\te}_9}{}^a$ is independent of $\eta$, it will not carry 
such an index. Let us also associate, in the obvious way, an index
$\eta$ to the vielbein with flat spinor indices.

%%%%%%%%%%%%%%%%%%%%%%%%%%%%%%%%%%%%%%%%%%%%
%%%%%%%%%%%

Following the strategy explained above,
let us parametrize the dual vielbein via their projection
on the original vielbein as follows:
\bea
{E_{I\alpha}}{}^{\hat M}{{\tp_{(\eta)}}_{\hat M}}{}^{J\beta}&=&
{{{\cal M}_{(\eta)}}_{I\alpha}}{}^{J\beta}\nonumber\\
{E_{a}}{}^{\hat M}{{\tp_{(\eta)}}_{\hat M}}{}^{J\beta}&=&
{{{\cal A}_{(\eta)}}_a}{}^{J\beta}-
\eta_{ab}{{E_9}{}^b\ov G_{99}} 
{E_9}{}^{I\alpha}{{{\cal M}_{(\eta)}}_{I\alpha}}{}^{J\beta}\,\,\,\,
{\rm with}\,\,\,\,
{E_9}{}^a{{{\cal A}_{(\eta)}}_{a}}{}^{I\beta}=0\nonumber\\
{{\te_{(\eta)}}_9}{}^{I\alpha}&=&{a_{(\eta)}}{}^{I\alpha}(E)
\label{eq:1}
\eea
where ${\cal M}{}^{IJ}$, ${\cal A}$ and $a$ are for the time being 
arbitrary functions of the IIA fields. The chirality of the 
original vielbein implies that 
${(\Gamma_+)_\alpha}{}^\beta{{\cal M}_{1\beta}}{}^{J\gamma}=0$ 
and
${(\Gamma_-)_\alpha}{}^\beta{{\cal M}_{2\beta}}{}^{J\gamma}=0$. 
Equations (\ref{eq:1}) completely describe ${{\te_{(\eta)}}_M}{}^{I\alpha}$
as can be seen from the following counting of degrees of freedom:
for a given $\eta$,
${\cal M}_{(\eta)}$ with its chirality constraints accounts for 
$2\times 2\times{1\ov 2}\times 32\times 32$ independent components
while ${\cal A}$ and $a$ account for $9\times 2\times 32$ 
and $2\times 32$, respectively. We have therefore the number of 
independent components of ${{\te_{(\eta)}}_M}{}^{I\alpha}$.  On the other hand,
the duality transformation does not change the number of independent 
degrees of freedom. The chirality constraints on ${\cal M}$ require
that its rank is at most $32$ which is also the number of degrees of 
freedom in the original theory. Therefore, we expect that
all field redefinitions will involve only non-degenerate matrices.

We can easily invert the equations (\ref{eq:1}) and express
${{\tp_{(\eta)}}_{\hat M}}{}^{I\alpha}$ in terms of IIA fields and 
the arbitrary objects ${\cal M}$ and ${\cal A}$.
Introducing the notation  ${{P}_{\hat M}}{}^{IA}=
{E_{\hat M}}{}^{IA}-{G_{{\hat M}9}\ov G_{99}}{E_{9}}{}^{IA}$, 
the solution reads:
\bea
{{\tp_{(\eta)}}_{\hat M}}{}^{I\alpha} &=&{{P}_{\hat M}}{}^{J\beta}
{{{\cal M}_{(\eta)}}_{J\beta}}{}^{I\alpha}+{{P}_{\hat M}}{}^{b} 
{{{\cal A}_{(\eta)}}_b}{}^{I\alpha}\nonumber\\
{{{\te}_{(\eta)}}_{9}}{}^{I\alpha} &=& a_{(\eta)}{}^{I\alpha}(E)
\label{eq:solspinor}
\eea
where we have also used that $ {E_9}{}^a{{\cal A}_a}{}^{I\beta}=0$.
This is the most general solution since 
equations (\ref{eq:1}) form a linear system with as many equations 
as unknowns.
At this point it seems natural to introduce also the object
${{\tp_{(\eta)}}_{\hat M}}{}^{a}={{\te}_{(\eta) \hat M}}{}^a-
{{\tilde G}_{{\hat M}9}\ov {\tilde G}_{99}}{{\te}_{(\eta)9}}{}^a$ and notice 
that it is equal to ${{P}_{\hat M}}{}^{a}$ upon using the explicit 
form (\ref{eq:solb}) of vielbein with flat bosonic index.
All these can be summarized in the following equation:
\be
\pmatrix{{{\tp_{(\eta)}}_{\hat M}}{}^{a}&{{\tp_{(\eta)}}_{\hat M}}{}^{I\alpha}
\cr}=
\pmatrix{{{P}_{\hat M}}{}^{b}& {P_{\hat M}}{}^{J\beta}\cr}
\pmatrix{\delta_b{}^a&{{{\cal A}_{(\eta)}}_b}{}^{I\alpha}\cr 0&
{{{\cal M}_{(\eta)}}_{J\beta}}{}^{I\alpha}\cr}
\label{eq:fff}
\ee 
In the following we will enforce the chirality constraints 
on ${\cal M}$ by expressing it as appropriate projectors acting on 
nondegenerate matrices.

Acting on equation (\ref{eq:fff}) from the right with
matrices of the type $\pmatrix{\id&C\cr 0&\id\cr}$ we notice that
the value of ${\cal A}$ can be changed arbitrarily. At the same time,
such transformations are nondegenerate for any $C$ 
and leave the Lagrangian 
invariant. They are, therefore, perfectly valid field redefinitions.
Using them we will set ${\cal A}=0$ in the following. 
Notice that in this procedure we did not use the extra constraints 
coming from Siegel symmetry.

By requiring that equation(\ref{eq:second}) is covariant we see that
the matrices ${\cal M}$ must be of the form:
\be
{{{\cal M}_{(\eta)}}_{I\alpha}}{}^{J\beta}=
{{{\cal T}_{(\eta)}}_{I\alpha}}{}^{K\gamma}\delta_K{}^L
{(\Gamma_{-\eta})_\gamma}{}^\rho
{{{\cal N}_{(\eta)}}_{L\rho}}{}^{J\beta}
\label{eq:cov}
\ee
where ${\cal N}$ are arbitrary, covariant, nondegenerate 
objects, ${\cal T}$ was defined in equation
(\ref{eq:deft}) and $\Gamma_{-\eta}$ is defined to be $\Gamma_{\pm}$
for $\eta=\mp$. We have also explicitely 
taken into account the chirality 
constraints by inserting the appropriate projectors.
We could of course ask ourselves whether there really exists a
nontrivial ${\cal N}$ constructed out of covariant objects in the theory 
which also has the right dimensions. As we will see below, there is
no need to construct the most general such object since
by knowing that it is nondegenerate we can absorb it by redefining
${{\te}_M}{}^{I\alpha}$. This will completely fix the Siegel symmetry 
transformations. In the following we will construct the inverse
of the matrix ${\cal N}$.

Requiring that T-duality is an involution provides additional 
informations regarding ${\cal M}_{(\eta)}$ and $a_{(\eta)}$. Following the 
same steps as before but this time using (\ref{eq:second}) for 
the Siegel transformations, we get the following equation:
\be
{{{\cal T}_{(\tilde \eta)}}_{I\alpha}}{}^{K\gamma}
{{{\cal T}_{(\eta)}}_{K\gamma}}{}^{P\epsilon}
{{{\cal M}_{(\eta)}}_{P\epsilon}}{}^{L\rho}
{{\tcm_{(\eta,{\tilde\eta})}}_{L\rho}}{}^{J\beta}=
\big(\delta_{|\eta-{\tilde\eta}|, 
|I-J|}\big)_I{}^J\big[{(\Gamma_-)_\alpha}{}^\beta\delta_J{}^1+
{(\Gamma_+)_\alpha}{}^\beta\delta_J{}^2 \big]
\label{eq:invol}
\ee  
where ${{{\tcm}_{(\eta,{\tilde\eta})}}_{I\alpha}}{}^{J\beta}=
{{{\te}_{(\eta)}}_{I\alpha}}{}^{\hat M}
\big({{E_{({\tilde\eta})}}_{\hat M}}{}^{J\beta}-{G_{{\hat M}9}\ov
G_{99}}{{E_{({\tilde\eta})}}_9}{}^{J\beta}\big)$,
${{E_{({\tilde\eta})}}_M}{}^{I\beta}
={E_M}{}^{I\beta}$ if ${\tilde \eta}=\eta$
and ${{E_{({\tilde\eta})}}_M}{}^{I\beta}$ for ${\tilde \eta}=-\eta$ will be 
constructed shortly.
One might wonder why we are repeating all the computation and do not 
formally dualize equation (\ref{eq:fff}). The point is that there 
is additional information, related to covariance requirement, that 
has to be taken into account. 

We claim that it is enough to consider, say, ${\tilde \eta}=\eta$.
In order to prove this claim we notice
that $({\cal T}_{(\eta)}){}^2=\id\otimes\id$ 
and ${\cal T}_{(\eta)} {\cal T}_{(-\eta)}=\id\otimes{E\llap/}_9$.
We can therefore construct a map between ${\tilde \eta}=-\eta$ 
and ${\tilde \eta}=\eta$ theories: 
\be
{{{\tcm}_{(\eta,\eta)}{}^{J1}}}={{{\tcm}_{(\eta,-\eta)}{}^{J2}}}
{E\llap/}_9\,\,\,\,\,\,\,\,,
\,\,\,\,\,\,\,\,
{{{\tcm}_{(\eta,\eta)}{}^{J2}}}=
{{{\tcm}_{(\eta,-\eta)}{}^{J1}}}{E\llap/}_9\,\,\,\,.
\label{eq:obs}
\ee
This immediately translates into a map between the vielbein 
associated with the ${\tilde\eta}=\pm\eta$:
\be
{{E_{({\tilde\eta}=-\eta)}}_M}{}^{I\alpha}=
{{E_{({\tilde\eta}=\eta)}}_M}{}^{J\beta}|\epsilon{}^{JI}|
{(E\llap/_9)_\beta}{}^\alpha
\ee
Of course, this should be combined with a change in sign of the 
$X{}^9$ coordinate. At the level of Siegel transformations
this translates into the statement that 
$\delta_\kappa 
{E_{({\tilde\eta}=-\eta)}}{}^{I\alpha}\sim |\epsilon{}^{IJ}|\kappa{}^{J\alpha}$. 
Since it seems natural for the index on $E$ to match  the index on $\kappa$,
we relabel
the vielbein to absorb the $\epsilon$ symbol. The relabeled
vielbein have the chirality opposite to those obtained
by choosing ${\tilde \eta}=\eta$. We therefore conclude
that the two IIA theories are mapped into each other
by multiplying with $E\llap/_9$ the vielbein with flat spinor indices
and relabeling the result together with performing a parity 
transformation along the ninth coordinate. The last statement is subject 
to interpretation since the two IIA theories are actually identical.
From this point of view, the map described above is just an automorphism 
of the IIA theory.

In the following we will concentrate on the choice
${\tilde \eta}=\eta$. Combining equation (\ref{eq:invol})
with equation (\ref{eq:cov}) we see
that, since ${\cal N}{}^{IJ}$ is covariant, we
need ${{{\tcm}_{(\eta)}{}^{IJ}}}$ 
to be proportional to ${\tilde T}_{(\eta)}$. Thus, 
we are lead to rewrite $\tcm$ as follows:
$$
{{{\tcm}_{(\eta,\eta)}}_I}{}^J={{\tcn_{(\eta,\eta)}}_I}{}^K   
\Gamma_{-\eta}{{{\cal T}_{(\eta)}}_K}{}^J
$$
where ${\tilde{\cal N}}$ are, for now, arbitrary matrices and we have
taken care of chirality constraints.

Using this expression for ${\tilde{\cal M}}$, the
equation (\ref{eq:invol}) can be written as:
\be
\Gamma_{-\eta}{\cal N}_{(\eta)}{\tilde {\cal N}}_{(\eta,\eta)}\Gamma_{-\eta}
=\id\otimes\Gamma_{-\eta}
\ee
This shows that ${{{\tcn}_{(\eta,\eta)}}_I}{}^J$ is the 
inverse of ${{{\cal N}_{(\eta)}}_I}{}^J$ 
in the $\Gamma_{-\eta}$ sector and therefore 
enables us to redefine the vielbein
and completely absorb the arbitrariness in ${\cal N}{}^{IJ}$.
With the vielbein redefinitions
\be
{{{\te}_{(\eta)}}_M}{}^{I\alpha}\longrightarrow
{{{\te'}_{(\eta)}}_M}{}^{I\alpha}={{{\te}_{(\eta)}}_M}{}^{J\beta}
{{{\tcn}_{(\eta,\eta)}}_{J\beta}}{}^{I\alpha}
\label{eq:redef}
\ee
we therefore get the following dual fields:
\be
{\tp_{(\eta)}}{}^{I\alpha}={P}{}^{J\beta}
{{{\cal T}_{(\eta)}}_{J\beta}}{}^{I\gamma}
{(\Gamma_{-\eta})_\gamma}{}^\alpha
\label{eq:pp}
\ee
or, explicitly,
\be
\te_{(\eta)}{}_{\hat M}{}^{I\alpha}=(E_{\hat M}{}^{J\beta}
-{G_{{\hat M}9}\ov G_{99}}E_9{}^{J\beta}){\cal T}_{(\eta)}
{}_{J\beta}{}^{I\gamma}{(\Gamma_{-\eta})_\gamma}{}^\alpha+
\eta B_{{\hat M}9}\te_{(\eta)}{}_{9}{}^{I\alpha}
\label{eq:explic}
\ee
Thus,
we see that all arbitrariness of ${\cal M}$ and ${\cal A}$
can be removed by suitable field redefinitions. The left over
piece in equation (\ref{eq:pp}) (${{{\cal T}_{(\eta)}}_{J\beta}}
{}^{I\gamma}$) is needed to restore 
the covariance of the dual Siegel transformations.
$\te_{(\eta)}{}_{9}{}^{I\alpha}$ is left arbitrary by the above analysis,
but it will be fixed shortly from other considerations. Its chirality
is, however, fixed by equation (\ref{eq:pp}).
Of course, the chiral projector can be absorbed by 
pushing it past ${\cal T}$ and letting it act on $P{}^{I\alpha}$.
The Siegel transformations in the dual theory take the following
form:
\be
\delta_\kappa {\te_{(\eta)}}{}^{I\alpha} =
\partial_i{\tilde Z}{}^M{{{\te}_{(\eta)}}_M}{}^a
P_{\cal I}{}^{ij} {{\tilde\kappa}_{(\eta) j}}{}^{I\beta}
{(\Gamma_a\Gamma_{-\eta})_\beta}{}^\alpha\,\,\,\,\,\,\,\,\,\,\,\,
{\cal I}=\left\{\begin{array}{rrc}
+\,\,\,{\rm if}\,\,\,I=1\\
-\,\,\,{\rm if}\,\,\,I=2
\end{array}
\right.
\label{eq:vc1}
\ee
\noindent
From equation (\ref{eq:pp}) we also see that both types of 
vielbein in the dual theory have the same chirality.
Let us note that by redefining 
${\te}{}^{\alpha,\bar{\alpha}} \longleftrightarrow
 {\te}{}^{1\alpha} \mp i{\te}{}^{2\alpha}$
one recovers\footnote{In reference \cite{GHMNT} the factor of $i$ in the
above redefinition is absorbed in the spinor metric} the Siegel symmetry 
transformations
of type IIB superstring theory as stated in \cite{GHMNT}.

All possible choices of bosonic 
vielbein can be summarized in the following diagram:
\begin{figure}[hbtp]
\begin{center}
\mbox{\epsfxsize=9truecm\epsfysize=4truecm
\epsffile{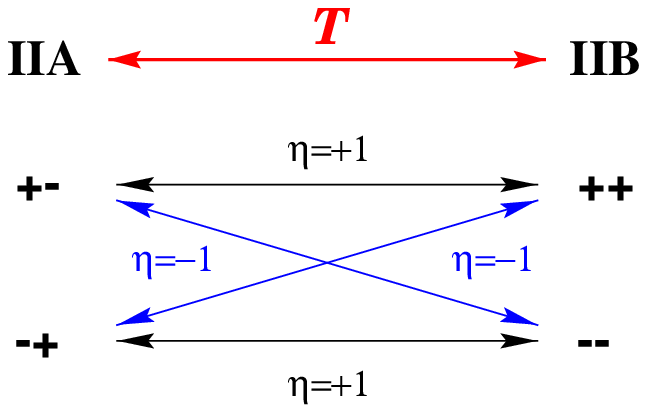}}
\end{center}
\centerline{{\bf Fig. 1.} T-dualities of type II string theories.}
\end{figure}

\noindent
The $+$ and $-$ signs in the figure represent the chiralities
of the two space-time supercharges, in the order $(Q{}^1,\,Q{}^2)$.

In order to fully determine the Buscher rules at the level 
of supervielbein we must now
analyze the situation of ${{\te_{(\eta)}}_9}{}^{I\alpha}$. Since the 
Siegel transformations depend only on ${{\tp_{(\eta)}}_{\hat M}}{}^{I\alpha}$
they cannot fix the functional form of $a{}^{I\alpha}_{(\eta)} (E)$. On the 
other hand $a{}^{I\alpha}_{(\eta)}$ must have the same chirality as 
${{\tp_{(\eta)}}_{\hat M}}{}^{I\alpha}$ since the latter contains
${{\te_{(\eta)}}_9}{}^{I\alpha}$ in its definition. Also, for \
${\tilde\eta}=\eta$, the involution requirement implies the following:
\be
{{\te{}^I_{({\tilde \eta})}}_9}{}^\alpha=
{{\tilde a}_{({\tilde \eta})}}{}^{I\alpha}(E)=
{a_{(\eta)}}{}^{I\alpha}({\te_{\hat M}}{}^B, {\te_9}{}^b,  
{{\te}_9}{}^{J\beta})=
{a_{(\eta)}}{}^{I\alpha}({\te_{\hat M}}{}^A, {\te_9}{}^a, 
a_{(\eta)}{}^{J\beta}(E))
\label{eq:arb}
\ee
This equation has many independent solutions.
At the same time, the superspace diffeomorphism covariance implies 
that $a{}^\alpha$ depends 
only on ${E_9}{}^A$ since the parameters do not depend on the ninth bosonic
coordinate and ${\te_9}{}^A$ should transform as the ninth component
of a covariant vector. The index structure and chirality requirements
for ${\te_9}{}^{I\alpha}$ suggest that
\be
{{\te_{(\eta)}}_9}{}^{I\alpha}= {{E_9}{}^{J\beta}\ov f_{(\eta)J}(G_{99})} 
{{{\cal T}_{(\eta)}}_{J\beta}}{}^{I\alpha}\,\,\,\,.
\label{eq:undetf}
\ee
From equation (\ref{eq:arb}) we see that the function $f_{(\eta)J}$ has to 
be a monomial, but its exact expression is arbitrary. To completely fix the 
form of $f_{(\eta)J}$
we study in detail the supergravity constraints. We will require that
the IIB supergravity constraints are satisfied if the IIA constraints are.
As pointed out before, this is not an outside constraint, since the Siegel symmetry, 
having the right form, already produces the right conventional constraints
as well as the other lower dimensional constraints, while the Bianchi identities 
are responsible for the rest.

\newsection{Matching the supergravity constraints}

In this section we will study the change in supergravity constraints 
under T-duality. 
We will first fix the arbitrariness in the functions $f_{(\eta)I}(G_{99})$ 
by comparing  
the constraints computed with our dual fields (equations (\ref{eq:BMN}), 
(\ref{eq:pp}) and 
(\ref{eq:undetf})) with those given in \cite{gates}. Then we compute the 
torsion constraints.
As a nice consistency check we will notice that the dimension $0$ 
constraints are satisfied
if the IIA constraints are satisfied. From one of the consequences of 
Siegel symmetry 
we infer the dual spin connections. Then, the dimension $1/2$ constraints 
will produce for
us the duality transformation of the dilaton and its superpartner. The 
constraints of
dimension $1$ will generate the transformations of RR fields.

It turns out that the easiest way  to fix $f_{(\eta)I}(G_{99})$ is to 
study the dual NS-NS field
strength. We will keep $\eta$ arbitrary for the time being. Later on we 
will fix it to $\eta=1$.
To flatten the indices we need the inverse of the dual vielbein. It is given by:
\bea
{\te_a}{}^{\hat M}\!\!\!\! &=&\!\!\!\! {E_a}{}^{\hat M} - \e9{}_a
\,{E_9}{}^{\alpha I}\,{E_{\alpha I}}{}^{\hat M} 
s_{(\eta)I} \nonumber\\
{\te_a}{}^9 \!\!\!\!&=&\!\!\!\! \e9{}_a
\LB
1 + s_{(\eta)I}
\,{E_9}{}^{\alpha I}\,{E_{\alpha I}}{}^{\hat M} B_{{\hat M}9}
\RB - {E_a}{}^{\hat M} B_{{\hat M}9} \\
{\te_{\alpha I}}{}^{\hat M} \!\!\!\!
&=& \!\!\!\!{\t_{\alpha I}}{}^{\beta J} {E_{\beta J}}{}^{\hat M} ~~~~;~~~~
{\te_{\alpha I}}{}^9 = -\,{\t_{\alpha I}}{}^{\beta J} {E_{\beta J}}{}^{\hat M}\,B_{{\hat M}9}
\nonumber
\eea
Here summation over the index $I$ is assumed and 
$s_{(\eta)I} = {1\over f_{(\eta)I}(G_{99})} - {1\over G_{99}}$. Using this and equation (\ref{eq:atf})
we get the following for the dual 3-form field strength:
\begin{eqnarray}
\tf_{ABC} =F_{A'B'C'} {\t_A}{}^{A'} {\t_B}{}^{B'} {\t_C}{}^{C'}-
(ABC)-(-)^{A(B+C)}(BCA)-(-)^{C(A+B)}(CAB)
\label{eq:fsm}
\end{eqnarray}
with $(ABC)$ given by 
\bea
(ABC)&=&{E_{9a}}\delta^a_{C}\,\Big\{ 
{1\over G_{99}}  \e9{}^d F_{d A' B'} {\t_A}{}^{A'} {\t_B}{}^{B'} 
+{1\ov f_{(\eta)I}(G_{99})} \e9{}^{\alpha I} 
F_{\alpha I A'B'} {\t_A}{}^{A'} {\t_B}{}^{B'}\label{eq:fs}\\
&+&\!\!\!\! {\eta\over G_{99}} {\T_{A' B'}}{}^a \e9{}_a 
{\t_A}{}^{A'} {\t_B}{}^{B'} 
+ {\eta\over G_{99}}  \e9{}^D {\T_{D B' d}}\delta_A^d 
{\t_B}{}^{B'} - {\eta\over G_{99}}  \e9{}^D {\T_{D A' d}}\delta_B^d 
{\t_A}{}^{A'} \Big\}~.\nonumber
\eea
with 
${\T_{AB}}^C={T_{AB}}^C+{E_A}^N{\Omega_{NB}}^C - 
(-)^{AB}{E_B}^N{\Omega_{NA}}^C$.

The 3-form field strength constraints summarized in 
the Appendix set to zero the 
components ${\tilde F}_{I\alpha\,J\beta\,K\gamma}$ in both IIA and IIB theories. 
It is easy to check that this is trivially satisfied, since 
${E_{9a}}\delta^a_{A}=0$ if $A$ is not a bosonic index.  

By considering the constraints on ${\tilde F}_{I\alpha\,J\beta\,c}$ we do not 
get any information
about the form of $f_{(\eta) I}$, but we have a nontrivial consistency check for 
the results obtained from Siegel symmetry arguments.

In both IIA and IIB supergravity one requires that the field strength 
components with two bosonic and one fermionic indices vanishes. Imposing 
this on equations (\ref{eq:fsm}) and (\ref{eq:fs}) amounts to canceling
${\te_{(\eta)9}}{}^{K\rho} F_{K\rho\, I\gamma\, a}$ against 
${\eta\ov G_{99}}{E_9}{}^{K\rho} {T_{K\rho\,I\gamma\,a}}$.
Using the IIA constraints we find that the functions $f_{(\eta) I}$ are given by
\be
f_{(\eta)1}=-\eta\, G_{99}
\,\,\,\,\,\,\,\,
\,\,\,\,\,\,\,\,\,\,\,\,f_{(\eta)2}=\eta\, G_{99}\,\,,
\ee
which together with equation (\ref{eq:explic}) recovers the results stated in 
equation (\ref{eq:finalrez}-\ref{eq:finalrez1}). 
This and the Siegel symmetry imply that all the constraints of IIB supergravity 
are satisfied. 
We are therefore free to use them to derive the T-duality action on fields that 
do not appear explicitly in the sigma model action.

%%%%%%%%%%%%%%%%%%%%%%%%%%%%%%%%%%%%%%%%%%%%%%%%%%%

Now we turn to studying the torsion constraints. Since the formulas are cumbersome enough,
we will set $\eta=1$. The $\eta=-1$ case can be recovered by interchanging the labels $1$ and $2$.
We will also write $s_I$ for $s_{(1)I}$. For notational convenience we
extend the index of $f$ to a full superspace index and define $f_a=G_{99}$.
Likewise, the index of $s$ is extended to a full superspace index and
$s_a=0$.  
First we write down the dual torsion in terms of the IIA variables. The dimension $0$ 
components will provide additional consistency checks to our previous results, the dimension 
$1/2$ components will determine the dual spin $1/2$ fields as well the dimension $1/2$ 
dual spin connection, while the dimension $1$ components will generate the rest of the dual spin 
connection components as well as the duality transformation of the RR fields.

Using the definitions and the constraints listed in the appendix we find the following for the
dual torsion:
\begin{eqnarray}
{{\tilde{\T}}_{AB}}{}^C &=&  {\T_{A'B'}}{}^{C'} {\t_A}{}^{A'} 
{\t_B}{}^{B'} {\t_{C'}}{}^{C}
-\Big\{  {1\over 2 f_C} \e9{}^D F_{DA'B'} \e9{}^C{\t_A}{}^{A'} 
{\t_B}{}^{B'} {\t_C}{}^{C'} 
\nonumber\\
&-& {s_{I}\over f_C} 
\e9_d \delta^d_A \e9{}^{I\alpha} \e9{}^D F_{DI\alpha B'} \e9{}^{C'}
{\t_B}{}^{B'} {\t_{C'}}{}^{C} \nonumber\\
&+& {1\over G_{99}} \e9{}^D {\T_{D B' d}}\delta_A^d \e9{}^{C'}
{\t_B}{}^{B'} {\t_{C'}}{}^{C} \nonumber\\
&+&{1\over 2G_{99}} {\T_{A' B'}}{}^a \e9_a \e9{}^{C'}
{\t_A}{}^{A'} {\t_B}{}^{B'} {\t_{C'}}{}^{C} \nonumber\\
&+& s_C \, \e9{}^D  {\T_{DA'}}{}^{C'} (\e9)_d \delta^d_B
{\t_A}{}^{A'}{\t_{C'}}{}^{C} \nonumber\\
&-&{s_{I} \over G_{99}} 
\e9_d\delta^d_A \e9{}^{I\alpha} {\T_{I\alpha B'}}{}^b \e9_b\e9{}^{C'} 
{\t_B}{}^{B'} {\t_{C'}}{}^{C} \nonumber\\
&-& {s_{I} \over G_{99}} 
\e9{}^{I\alpha} \e9{}^D {\T_{DI\alpha d}}\delta^d_A  \e9_d \delta_B^d
\e9{}^{C'}{\t_{C'}}{}^{C} \nonumber\\
&+& \partial_{A'} ( {1\over f_C} {\t_{C'}}{}^{C} - {1 \over G_{99}} )
\e9_d\delta_B^d \e9{}^{C'} {\t_A}{}^{A'}\nonumber\\
&+& s_{I } \,
\e9_d\delta_A^d \e9{}^{I\alpha} {\T_{I\alpha B'}}{}^{C'}
{\t_B}{}^{B'} {\t_{C'}}{}^{C} \nonumber\\
&+& {1\ov \sqrt{G_{99}}}
\e9{}^D {\T_{DA'}}{}^a {\delta_{B'}}{}^{2\alpha } 
(\Gamma_a\Gamma_+)_\alpha{}^\beta
{\delta_{2\beta }}{}^C 
{\t_A}{}^{A'} {\t_B}{}^{B'} \nonumber\\
&-& {1 \over G_{99}} \e9{}^D {\T_{DA'}}{}^a (\e9)_a {\delta_{B}}{}^{2\alpha }
{\delta_{2\alpha }}{}^{C} {\t_A}{}^{A'} \nonumber\\
&-& {1\ov \sqrt{G_{99}}} {1\over f_{2\alpha }} 
\e9{}^D {\T_{DA'}}{}^a \e9{}^{2\alpha } 
(\Gamma_a\Gamma_+)_\alpha{}^\beta
{\delta_{2\beta }}{}^C \e9_d{\delta_B^d} {\t_A}{}^{A'} \nonumber\\
&+& {1 \over G_{99}}{1\over f_{2\alpha }}  
\e9{}^D {\T_{DA'}}{}^a \e9_a \e9_d \delta_B^d
\e9{}^{2\alpha } {\es_{\alpha }}{}^{\beta }
{\delta_{2\beta }}{}^C {\t_A}{}^{A'} \nonumber\\
&-& { s_{I}\ov \sqrt{G_{99}}}  
\e9{}^{\gamma I} \e9{}^D {\T_{D\gamma I}}{}^a 
{\delta_{B'}}{}^{2\alpha }
(\Gamma_a\Gamma_+)_\alpha{}^\beta
{\delta_{2\beta }}{}^C 
\e9_d{\delta_A^d} {\t_B}{}^{B'} \nonumber\\
&+& { s_{I}\ov G_{99}}  \e9{}^{\gamma I} \e9{}^D {\T_{D\gamma I}}{}^a 
\e9_a \e9_d{\delta_A^d} {\delta_{B}}{}^{2\alpha }
{\delta_{2\alpha }}{}^{C} \Big\} +(-)^{AB}(BA)
\label{eq:torsion}
\end{eqnarray}
As usual, the spin connection with fermionic indices (which appear in $\tilde{\T}$ and $\T$) 
is related to the one with bosonic Lorentz indices by
\be
{\Omega_{A \alpha}}{}^\beta = 
{1\ov 8}{\Omega_{A, a}}{}^b{({\sigma{}^a}_b)_\alpha}{}^\beta
\ee
where, as stated in the Appendix, $\sigma{}^{a b}$ contains no factor of $1/2$.

As promised earlier, equation (\ref{eq:torsion}) will provide the T-duality transformations 
of the remaining (super)fields
as well as more consistency checks. Starting with torsion components of dimension
$0$ and writing the result in 16-component notation, we find that 
\be
{\tt_{1\alpha\,2\beta}}^a=0~~~;~~~{\tt_{1\alpha\,1\beta}}^a=i(\sigma^a)_{\alpha\beta}~~~;~~~
{\tt_{2\alpha\,2\beta}}^a=i(\sigma^a)_{\alpha\beta}~.
\ee
These are indeed the correct constraints of IIB supergravity as shown in \cite{gates}
and summarized in the Appendix.

We proceed by finding the relation between the spin connections in the type II theories.
They arise from constraints of dimension $1/2$ and $1$. The fact that the 
Siegel symmetry has the correct form implies that
${\tt_{{a} b}}{}^c = 0 $ and ${\tt_{{I\alpha} b}}{}^c = 0 $. Using this information 
in equation  (\ref{eq:torsion})  we find that 
\be
{\to_{1\alpha  b}}{}^a = {\o_{1\alpha  b}}{}^a
+ is_1\e9{}^{1\beta } \Big( (\Gamma^a\Gamma_+ C^{-1})_{\beta\alpha}
\e9_b - 
(\Gamma_b\Gamma_+ C^{-1})_{\beta\alpha} \e9{}^a 
\Big) ~~~~~
{\to_{2\alpha  b}}{}^a = {\es_{\alpha }}{}^{\beta } {\o_{2\beta  b}}{}^a
\ee
and 
\bea
{\to_{[ab]}}{}^c = {\o_{[ab]}}{}^c 
- {1 \over G_{99}} \e9{}^d \o_{[db]a} \e9{}^c 
+ {1 \over G_{99}} \e9{}^d \o_{[da]b} \e9{}^c 
- {1 \over G_{99}} \o_{[ab]d}  \e9{}^d  \e9{}^c \\
- {1 \over G_{99}}  \e9{}^d F_{dab} \e9{}^c
- {2 \over G_{99}} \e9{}^{I\alpha} \o_{{I\alpha}ba} \e9{}^c
+ {2 \over G_{99}} \e9{}^{1\alpha } \e9_{[a|} {\o_{1\alpha |b]}}{}^c~.
\eea
From the last equation follows ${\o_{ab}}{}^c$ as
\be
{\o_{ab}}{}^c = {1 \over 2} ( {\o_{[ab]}}{}^c + {\o_{[ca]}}{}^b - {\o_{[bc]}}{}^a )~~.
\label{eq:connection}
\ee

Knowing the spin connections it is now easy to proceed and find the dual spin $1/2$ fields.
They come from constraints of dimension $1/2$.
From ${\tt_{1\alpha  1\beta }}{}^{\gamma 1} $  and ${\tt_{2\alpha  2\beta }}{}^{\gamma 2}$ 
follows that  $\tl_{I\alpha}$ are given by:
\bea
2\tl_{1\alpha } =& \Lambda_{1\alpha } - \partial_{1\alpha } \ln (G_{99}){}^{-1/2} 
&= \Lambda_{1\alpha } + i(G_{99}){}^{-1/2} \e9{}^{\gamma 1}
(\es C^{-1})_{\gamma\alpha} \nonumber\\
2\tl_{2\alpha } =& {\es_\alpha}{}^{\dot\beta} ( 
\Lambda_{2\beta } - \partial_{2\beta } \ln (G_{99}){}^{-1/2} ) &
= (\es \Lambda_2)_\alpha - i(G_{99}){}^{-1/2} \e9{}^{\gamma 2} 
(C^{-1})_{\alpha{\dot\gamma}}
\label{eq:spin1/2}
\eea
\noindent
Since $ \tl_{\alpha} = {\te_{\alpha}}{}^M \partial_M \tph $ it follows that the
dual dilaton is given by
\be
2\tph = \ph - \ln (G_{99}){}^{-1/2}~.
\ee
This leads us to interpret $2\tph$ of reference \cite{gates} as the type IIB 
dilaton and
thus get the standard dilaton shift under T-duality, as shown in \cite{busher}. 
Opposite 
to that paper and as noted in \cite{CLPS}, in this treatment the dilaton shift 
appears
at the classical level. This can be intuitively understood from the fact that 
supergravity 
is already a one loop effect from the sigma model point of view.

The constraints of dimension $1$, besides determining the transformation rules
of RR fields, provide some more consistency checks. In particular, it can 
be shown that, upon using (\ref{eq:torsion}) together with
\bea
\tf_{abc} = {1\over G_{99}} \Big[ - \e9^d F_{dab} (\e9)_c
- {\o_{ab}}^d (\e9)_d (\e9)_c \nonumber\\
- (\e9)^D \o_{Dba} (\e9)_c
+ (\e9)^D \o_{Dab} (\e9)_c \Big]_{[abc]}
\label{eq:fieldstrengthabc}
\eea
and equation (\ref{eq:connection}), imply that the following equations are satisfied:
\be
\tt_{a1\alpha}{}^{1\beta}=-{1\ov 16}(\Gamma^{bc}\Gamma_-)_\alpha{}^\beta 
\tf_{abc}~~~~~~~~
\tt_{a2\alpha}{}^{2\beta}={1\ov 16}(\Gamma^{bc}\Gamma_-)_\alpha{}^\beta 
\tf_{abc}~~.
\ee
These are nothing but equations (\ref{eq:a20}) from the list of 
IIB constraints in the Appendix.

The RR (super)field transformations follow from the 
${\tt_{a1\alpha }}{}^{2\beta }$ and ${\tt_{a2\alpha }}{}^{1\beta }$ components
of equation (\ref{eq:torsion}). They are given by:

\noindent
-the axion
\bea
-i \nabla_b (W - {\bar W}) &=& {1\ov 2}\e9{}^c F_{cb}-{1\ov 4}
{\sl e}^\phi\e9{}^d\Lambda_{1\alpha} (C\Gamma_{db}\Gamma_+)^{\alpha\beta}
\Lambda_{2\beta}
+{7\ov 4}
{\sl e}^{2\tph}\tl_{1\alpha} (C\Gamma_{b}\Gamma_-)^{\alpha\beta}
\tl_{2\beta}\nonumber\\
+&{}&\!\!\!\!\!\!\!\!\!\!\!
{\sl e}^\phi{\e9{}_b\ov 2G_{99}}\e9{}^{1\alpha}(\Gamma_-C^{-1})_{\alpha\beta}
\e9{}^{2\beta}
-{\sl e}^\phi{\e9{}^d\ov 4G_{99}}\e9{}^{1\alpha}
(\Gamma_-\Gamma_{db}C^{-1})_{\alpha\beta}\e9{}^{2\beta}
\label{eq:ax}
\eea

\noindent
-the RR 3-form field strength
\bea
-i(G_{bcd}-{\bar G}_{bcd})  -i (W - {\bar W}) \tf_{bcd} &=& {1\ov 4}\e9{}_{[b}F_{cd]}
-\e9{}^a F'_{abcd}\nonumber\\
&-&{{\sl e}^\phi\ov 8 ~G_{99}}\e9_{[b}\e9{}^{1\alpha}
(\Gamma_-\Gamma_{cd]}C^{-1})_{\alpha\beta}\e9{}^{2\beta}\nonumber\\
&+&{{\sl e}^\phi\ov 2~4!~G_{99}}
\e9{}^a\e9{}^{1\alpha}(\Gamma_-\Gamma_{abcd}C^{-1})_{\alpha\beta}\e9{}^{2\beta}\nonumber\\
&-&{1\ov 8}{\sl e}^\phi\e9_{[b}\Lambda_{1\alpha}
(C\Gamma_{cd]}\Gamma_+)^{\alpha\beta}\Lambda_{2\beta}\nonumber\\
&-&{1\ov 2~4!}{\sl e}^\phi\e9^{a}\Lambda_{1\alpha}
(C\Gamma_{abcd}\Gamma_+)^{\alpha\beta}\Lambda_{2\beta}\nonumber\\
&+&{1\ov 3}{\sl e}^{2\tph}\tl_{1\alpha}
(C\Gamma_{bcd}\Gamma_-)^{\alpha\beta}\tl_{2\beta}
\label{eq:rr3}
\eea
which together with equation (\ref{eq:fieldstrengthabc})
fixes the imaginary part of $G_{abc}$, i.e. the RR 3-form field strength.

\noindent
-the RR 5-form field strength
\bea
\tf_{abcde} \!\!&=& \!\!{-1\ov 2~4!}\left[\e9_{[a} {F'}_{bcde]} + {1\over 5!} 
{\epsilon_{abcde}}{}^{fghij}\e9{}_{[f}{F'}_{ghij]}\right]-
{{\sl e}^{2\tph}\ov 8~5!}
\tl_{1\alpha}(C\Gamma_{abcde}\Gamma_-)^{\alpha\beta}\tl{}_{2\beta}\\
+~~~~\,&{}&\!\!\!\!\!\!\!\!\!\!\!\!\!\!\!\!\!\!\!\!\!\!\!
{{\sl e}^\phi\ov 4~(4!)^2~G_{99}}\left[\e9{}_{[a}\e9{}^{1\alpha}
(\Gamma_-\Gamma_{bcde]}C^{-1})_{\alpha\beta}\e9{}^{2\beta}+
{1\ov 5!}\epsilon_{abcde}{}^{ijklm}
\e9{}_{[i}\e9{}^{1\alpha}
(\Gamma_-\Gamma_{jklm]}C^{-1})_{\alpha\beta}\e9{}^{2\beta}\right]\nonumber\\
+&{}&\!\!\!\!\!\!\!\!\!\!\!\!
{{\sl e}^\phi\ov 4~(4!)^2}\left[\e9{}_{[a}\Lambda{}_{1\alpha}
(C\Gamma_{bcde]}\Gamma_+)^{\alpha\beta}\Lambda{}_{2\beta}+
{1\ov 5!}\epsilon_{abcde}{}^{ijklm}
\e9{}_{[i}\Lambda{}_{1\alpha}
(C\Gamma_{jklm]}\Gamma_+)^{\alpha\beta}\Lambda{}_{2\beta}\right]\nonumber
\label{eq:rr5}
\eea
This equation  provides another consistency check. By simple inspection one
can easily check that $\tf_{abcde}$ is self-dual, 
as it should be.

%%%%%%%%%%%%%%%%%%%%%%%%%%%%%%%%%%%%%%%%%%%%
%%%%%%%%%%%%%%%

\newsection{Conclusions}

This completes 
the construction of the T-duality action on the massless superfields of type II 
superstring theory, to all orders in the odd superspace coordinates. The results
are summarized in equations (\ref{eq:finalrez}-\ref{eq:finalrez1}) for 
the supervielbein and super 2-form, in equation (\ref{eq:spin1/2}) for spin 1/2
fermions and in equations (\ref{eq:ax}), (\ref{eq:rr3}) 
and (\ref{eq:rr5}) for the RR superfields.

It is easy to see that the dual gravitini obtained here 
agree with those derived in \cite{H1}. We have also constructed the
T-duality transformations of the RR field strengths by studying the 
constraints
of the type II supergravity theories. Another way of rederiving them 
would be to expand 
the supervielbein and super 2-form in the odd superspace coordinates.
The in principle, results should agree with those obtained in \cite{CLPS}, 
since 
the duality transformation commutes with the series expansion. However, possible
discrepancies can be associated to differences in the space-time interpretation 
of the fields.

The approach taken in this paper has the advantage of producing the
action of T-duality on all the space-time fermions without explicit reference
to space-time supersymmetry transformation rules and, at the same time, 
providing a compact form of all dual space-time fields.

An interesting exercise would be to relax the constraint that all fields
in the original theory are independent of $X{}^9$ by replacing it with
isometries characterized by some Killing vectors $k_M$. Such an approach
should confirm once again the $SO(d,d;\Z)$ \cite{RG} structure of T-duality for
IIA/B theory on a manifold with $d$ Killing vectors and should allow the 
construction of these transformations to all orders in the odd superspace 
variables.

\noindent
{\bf Acknowledgment}

\noindent We would like to thank Martin Ro\v cek for suggesting 
this problem and for usefull suggestions and discussions.
This work was supported in part by the NSF grant PHY-9722101.

\setcounter{section}{0}
\setcounter{subsection}{0}

\appendix{}

In this appendix we list the conventions we are using as well as the 
IIA and (appropriately redefined) IIB supergravity constraints. 

We use both  16 and 32-component notation. $\Gamma$ matrices are defined
in terms of two sets of generalized Pauli matrices 
${(\sigma_a)_\alpha}{}^{\dot\beta}$ and 
${({\bar\sigma}_a)_{\dot\alpha}}{}^{\beta}$
that have the following anticommutation relations:
\be
\sigma_{(a},{\bar \sigma}_{b)} = -2 \eta_{ab}
\ee
with the ``mostly plus'' metric.
The 16-dimensional indices are raised and lowered with the charge conjugation matrix 
$\C{}^{\alpha{\dot\beta}}$ and its inverse $\C_{\alpha{\dot\beta}}$:
\be
(\sigma_a)_{\alpha\beta} = {(\sigma_a)_\alpha}{}^{\dot\beta} 
\C_{\beta{\dot\beta}}~~~~~
({\bar\sigma}_a)_{{\dot\alpha}{\dot\beta}} =
{({\bar\sigma}_a)_{\dot\alpha}}{}^{\beta} \C_{\beta{\dot\beta}} ~~~~~\C^{\alpha{\dot\beta}}
\C_{\alpha{\dot\gamma}}=\delta^{\dot\beta}_{\dot\gamma}
\ee
The antisymmetric 
product of $n$ $\sigma$ matrices, $\sigma{}^{(n)}$ used here, and in general the 
antisymmetric product of $n$ objects, without  a factor of ${1\ov n!}$, e.g.
\be
{(\sigma_{ab})_\alpha}{}^\beta = -
{(\sigma_a)_\alpha}{}^{\dot\beta} 
{({\bar\sigma}_b)_{\dot\beta}}{}^{\beta} +
{(\sigma_b)_\alpha}{}^{\dot\beta} 
{({\bar\sigma}_a)_{\dot\beta}}{}^{\beta}
\ee

\noindent
$(\sigma_a)_{\alpha\beta}$, $(\sigma_{(5)})_{\alpha\beta}$ are symmetric in 
spinor indices while $(\sigma_{(3)})_{\alpha\beta}$ is antisymmetric. The barred
version of the matrices have the same symmetry properties as the unbarred ones 
and carry dotted indices. We will occasionally drop the bars in situations
where there is no possibility of confusion.

$\Gamma$ matrices are constructed as
\be
\Gamma_\alpha{}^\beta=\pmatrix{0&i{\bar\sigma}_{\dot\alpha}{}^{\beta}\cr
                i{\sigma}_{\alpha}{}^{\dot\beta}&0}
\ee
and satisfy the usual anticommutation relations 
$\{\Gamma^a,\,\Gamma^b\}=2\eta^{ab}$.

In 32-component notation one can choose from two possible charge 
conjugation matrices,
$C_+$ and $C_-$, which differ by a factor of $\Gamma_{11}$. We 
choose to work with 
$C\equiv C_+$, and the associated identities are:
\be
C^T=-C~~~~\Gamma_a^T=-C\Gamma_a C^{-1}~~~~C^\dagger=C^{-1}
\ee
One can go back and forth between 16 and 32-component notations by inserting the 
appropriate chiral projectors as well as charge conjugation matrices.

\noindent
The general expression for torsion and 2-form field strength are :
\bea
{T_{AB}}^C&=&(-)^{BN}{E_A}^N {E_B}^M (\partial_M {E_N}^C - (-)^{MN} \partial_N {E_M}^C)
-{E_A}^N{\Omega_{NB}}^C + (-)^{AB}{E_B}^N{\Omega_{NA}}^C \nonumber\\
F_{ABC} &=& -(-){}^{C(M+N)+BM}{E_A}{}^M {E_B}{}^N {E_C}{}^K 
(\partial_{K} B_{MN} \nonumber\\
&{}&~~~~~~~~~~~~~~~~~~~~~~~~~~~~~~~~~~+(-)^{K(M+N)}\partial_{M} B_{NK}
+(-)^{N(K+M)}\partial_{N} B_{KM})
\label{eq:atf}
\eea

With these conventions,
the IIA constraints are \cite{gates}:
\be
{T_{1\alpha  1\beta }}{}^a=i(\sigma{}^a)_{\alpha\beta}~~~~~~
{T_{2\alpha  2\beta }}{}^a=i
(\sigma{}^a)_{{\dot\alpha}{\dot\beta}}
~~~~~~{T_{1\alpha  2\beta }}{}^a=0
\ee
\be
{F_{a1\alpha  1\beta }}=-i(\sigma_a)_{\alpha\beta}~~~~~~
{F_{a 2\alpha  2\beta }}=i(\sigma{}^a)_{{\dot\alpha}{\dot\beta}}
~~~~~~{F_{a1\alpha  2\beta }}=0
\ee
\be
F_{a b\, I\gamma}=0~~~~~~~~{T_{I\alpha b}}{}^c=0
\ee
\be
{T_{1\alpha  1\beta }}{}^{\gamma 1}= 
\left[ {\delta_{(\alpha}}{}^{\gamma}
{\delta_{\beta )}}{}^\delta + 
(\sigma{}^a)_{\alpha\beta}(\sigma_a){}^{\gamma\delta} 
\right] \Lambda_{\delta 1}
\ee
\be
{T_{2\alpha  2\beta }}{}^{\gamma 2}= \left[ 
{\delta_{({\dot\alpha}}}{}^{{\dot\gamma}}
{\delta_{{\dot\beta} )}}{}^{\dot\delta} + 
(\sigma{}^a)_{{\dot\alpha}{\dot\beta}}
(\sigma_a){}^{{\dot\gamma}{\dot\delta}} 
\right] \Lambda_{\delta 2}
\ee
\be
{T_{1\alpha  1\beta }}{}^{\gamma 2} = \,
{T_{1\alpha  2\beta }}{}^{\gamma 1} = \,
{T_{1\alpha  2\beta }}{}^{\gamma 2} = \,
{T_{2\alpha  2\beta }}{}^{\gamma 1} = \, 0
\ee
\be
{T_{a1\beta }}{}^{1\alpha } = - 
{1\over 16} {(\sigma_{bc})_\beta}{}^\alpha F_{abc}
~~~~~~~{T_{a2\beta }}{}^{2\alpha } = 
{1\over 16} {(\sigma_{bc})_{\dot\beta}}{}^{\dot\alpha} F_{abc}
\ee
\be
{T_{a1\alpha }}{}^{2\beta } =  
- {i\over 16} {(\sigma_{a})}_{{\alpha}{\gamma}} 
\left[
{1\over 2}{(\sigma{}^{bc})}{}^{{\dot\beta}{\gamma}} 
B_{bc} - {1\over 12} {1\over 4 ! } 
{(\sigma{}^{bcde})}{}^{{\dot\beta}{\gamma}} D_{bcde} 
\right]
\ee
\be
{T_{a2\alpha }}{}^{1\beta } =  {i\over 16} {(\sigma_{a})}_{{\dot\alpha}{\dot\gamma}}
\left[ 
{1\over2}(\sigma{}^{bc}){}^{{\dot\gamma}{\beta}} B_{bc} - {1\over12} {1\over4!}
(\sigma{}^{bcde}){}^{{\dot\gamma}\beta} D_{bcde} 
\right]
\ee
\be
B_{bc} = {\sl e}{}^{-\ph} F_{bc} 
- {1\over2} \Lambda_{1\beta }
(\sigma_{bc}){}^{{\dot\alpha}{\beta}}\Lambda_{2\alpha }
\ee
\be
D_{bcde} = 2 {\sl e}{}^{-\ph} F'_{bcde} + 
{1\over4!} \Lambda_{1\beta }
(\sigma_{bcde}){}^{{\dot\alpha}{\beta}}\Lambda_{2\alpha }
\ee
\noindent

The IIB constraints are also listed in \cite{gates}, 
together with the appropriate
field redefinitions which transform them to string frame.  
Redefining the upper indices as 
$1\alpha  = {\alpha + {\bar \alpha} \ov \sqrt{2} }$, 
$2\alpha  = -i{\alpha - {\bar \alpha} \ov \sqrt{2} }$ 
and the lower indices as
$1\alpha  = {\alpha + {\bar \alpha} \ov \sqrt{2} }$, 
$2\alpha  = i{\alpha - {\bar \alpha} \ov \sqrt{2} }$ 
they are\footnote{The index redefinition is necessary for 
the Siegel transformations to have the form derived 
in section~{\bf 3}. The upper and lower indices are 
redefined in a different way for the scalar product to 
be preserved.}: 
\be
{T_{1\alpha  1\beta }}{}^a=i(\sigma{}^a)_{\alpha\beta}~~~~~~
{T_{2\alpha  2\beta }}{}^a= i(\sigma{}^a)_{\alpha\beta}
~~~~~~{T_{1\alpha  2\beta }}{}^a=0
\ee
\be
{F_{a1\alpha  1\beta }}=-i(\sigma_a)_{\alpha\beta}~~~~~~
{F_{a 2\alpha  2\beta }}= i(\sigma{}^a)_{\alpha\beta}
~~~~~~{F_{a1\alpha  2\beta }}=0
\ee
\be
F_{a b\, I\gamma}=0~~~~~~~~{T_{I\alpha b}}{}^c=0
\ee
\be
{T_{I\alpha \beta J}}{}^{\gamma K}= 
2\left[ 
{\delta_{(\alpha}}{}^{\gamma}{\delta_{\beta )}}
{}^\delta + (\sigma{}^a)_{\alpha\beta}
(\sigma_a){}^{\gamma\delta}
\right] 
\Lambda_{\delta I}~\delta_{I,J} \delta_{I,K}
\ee
\be
{T_{a1\beta }}{}^{1\alpha } = 
- {1\over 16} {(\sigma_{bc})_\beta}{}^\alpha F_{abc}~~~~~~~
{T_{a2\beta }}{}^{2\alpha } = {1\over 16} {(\sigma_{bc})_\beta}{}^\alpha F_{abc}
\label{eq:a20}
\ee
\begin{eqnarray}
{T_{a2\alpha }}{}^{1\beta } &=&  -
{i\over 4} {(\sigma_a\sigma{}^b)_\alpha}{}^\beta
\left[ 
{\sl e}{}^{-2\ph} \nabla_b (W - {\bar W}) + 
{7\over 4} \Lambda_{\gamma 2}(\sigma_{b}){}^{\gamma\delta}\Lambda_{\delta 1} 
\right] \\
&-& {1\over 4!\, 48} {(\sigma_{bcde})_\alpha}{}^\beta
\left[ 
{5\over 3} {\sl e}{}^{-2\ph} {F_a}{}^{bcde} + {i\over 5!\, 8} 
\Lambda_{\gamma 2}({\sigma_a}{}^{bcde}){}^{\gamma\delta}\Lambda_{\delta 1} 
\right] \nonumber\\
&+& {i\over 4!\, 6} {(\sigma_a\sigma{}^{bcd})_\alpha}{}^\beta
\left[ 
{\sl e}{}^{-2\ph} 
\left( {\bar G}_{bcd} - G_{bcd} - (W - {\bar W}) F_{bcd} \right) 
 + 
{1\over 3} \Lambda_{\gamma 2}(\sigma_{bcd}){}^{\gamma\delta}\Lambda_{\delta 1} 
\right] \nonumber
\end{eqnarray}

\begin{eqnarray}
{T_{a1\alpha }}{}^{2\beta } &=& {i\over 4} 
{(\sigma_a\sigma{}^b)_\alpha}{}^\beta
\left[ 
{\sl e}{}^{-2\ph} \nabla_b (W - {\bar W}) + {7\over 4} 
\Lambda_{\gamma 2}(\sigma_{b}){}^{\gamma\delta}
\Lambda_{\delta 1} 
\right] \\
&+& {1\over 4!\, 48} {(\sigma_{bcde})_\alpha}{}^\beta
\left[ 
{5\over 3} {\sl e}{}^{-2\ph} {F_a}{}^{bcde} + 
{i\over 5!\, 8} \Lambda_{\gamma 2}({\sigma_a}{}^{bcde}){}^{\gamma\delta}\Lambda_{\delta 1} 
\right] \nonumber\\
&+& {i\over 4!\, 6} {(\sigma_a\sigma{}^{bcd})_\alpha}{}^\beta
\left[ 
{\sl e}{}^{-2\ph} 
\left( {\bar G}_{bcd} - G_{bcd} - (W - {\bar W}) F_{bcd} \right) 
 + 
{1\over 3} \Lambda_{\gamma 2}(\sigma_{bcd}){}^{\gamma\delta}\Lambda_{\delta 1} 
\right] \nonumber
\end{eqnarray}

For both IIA and IIB theories, higher dimensional constraints can be constructed
by using the lower dimensional ones together with the Bianchi 
identities \cite{gates}, \cite{HW}.

\end{document}